\documentclass[
aps,%
12pt,%
final,%
notitlepage,%
oneside,%
onecolumn,%
nobibnotes,%
nofootinbib,%
superscriptaddress,%
noshowpacs,%
centertags]%
{revtex4}

\begin{document}

\title{Dynamics of Gaseous Disks in a Non-axisymmetric Dark Halo}

\author{\firstname{A. V.}~\surname{Khoperskov}}
\affiliation{%
Volgograd State University, Volgograd, Russia
}%
\author{\firstname{M. A.}~\surname{Eremin}}
\affiliation{%
Volgograd State University, Volgograd, Russia
}%

\author{\firstname{S. A.}~\surname{Khoperskov}}
\affiliation{%
Volgograd State University, Volgograd, Russia
}%
\author{\firstname{M. A.}~\surname{Butenko}}
\affiliation{%
Volgograd State University, Volgograd, Russia
}%
\author{\firstname{A. G.}~\surname{Morozov}}
\affiliation{%
Volgograd State University, Volgograd, Russia
}%


\begin{abstract}
The dynamics of a galactic disk in a non-axisymmetric (triaxial) dark halo is studied in detail using high-resolution, numerical, hydrodynamical models. A long-lived, two-armed spiral pattern is generated for a wide range of parameters. The spiral structure is global, and the number of turns can be two or three, depending on the model parameters. The morphology and kinematics of the spiral pattern are studied as functions of the halo and disk parameters. The spiral structure rotates slowly, and its angular velocity varies quasi-periodically. Models with differing relative halo masses, halo semi-axis ratios, distributions of matter in the disk, Mach numbers in the gaseous component, and angular rotational
velocities of their halos are considered.
\end{abstract}

\maketitle

\section{INTRODUCTION}

Galactic gaseous and stellar disks are axisymmetric
only in a first approximation. Spiral density waves,
a central bar, the non-axisymmetry (triaxial nature)
of a dark massive halo, satellite galaxies, or a triaxial
bulge can disturb the axial symmetry of the "external" gravitational potential in which the galactic disk
component is situated. Results of dynamic modeling
of specific disk galaxies, designed to agree with
photometric and kinematic data, indicate the need to
take into account a dark halo of mass $M_h$ (which
exceeds the mass of the disk component $M_d$ by a
factor of one to four [1--5]) inside the optical radius
of the stellar component $R$. The kinematics of objects
in the immediate vicinity of the MilkyWay--M31 pair
indicate that the total mass inside 0.96 Mpc is $M_T\sim 2 \times10^{12}M_{\odot}$ [6], and the ratio of the masses of the
MilkyWay and M31 is $M_{MW}/M_{M31} = 4/5$. Since the
mass of the disk component of our Galaxy probably
does not exceed $5\times 10^{10}M_\odot$ [7], $\sim 15$\% of the total
mass within 1 Mpc belongs to the disk subsystems.

There is much theoretical and observational evidence
indicating the absence of central symmetry in
the dark-mass distribution within the disk components
of S galaxies. The triaxial shape of the halo can
be characterized by three space scales: $a$, $b$, $c$.

Studies of galaxy formation based on cosmological
N-body models indicate triaxial mass distributions in
cold-dark-matter (CDM) halos, with the following
semi-axis ratios obtained in different studies: $s=c/a\sim 0.6-0.7$, $q=b/a\sim 0.8-0.95$  [8];
$\langle{s}\rangle\simeq 0.7$, $\langle{q}\rangle\simeq 0.8$ [9];  $\langle{q}\rangle\simeq 0.75$ [10]. At the same
time, triaxial parameters s and q depend on the halo
mass, redshift, and distance to the halo center. The
dark halo is more symmetrical near the center than
at the periphery. The distributions turn out more
symmetrical in hot-dark-mattermodels than in CDM
models~[9].

Observations of the Sagittarius tidal stream
(Sagittarius dSph) indicate a lack of central symmetry
of the halo, with the axial ratio $c/a\gee 0.8$~[11].
However, if this stream is dynamically young, then
$c/a\simeq 0.6$ is possible [12]. Arc-shaped stellar tidal
streams found in the Sloan Digital Sky Survey
(SDSS), apparently resulting from the absorption of
dwarf galaxies, seem to be promising indicators of a
dark halo. The Virgo over-density structure could be
related to such peculiarities in the distribution of the
observed matter in the halo [13]. Stellar tidal streams
have been discovered in M31 [14] and a number of
outer galaxies (e.g., NGC 5907 [15], NGC 3310 [16],
NGC 1055 (see http://www.imagingdeepsky.com)).
These streams are apparently connected with the
disruption of dwarf galaxies in the massive triaxial
halo.

Observations of stars (RR Lyr, F dwarfs) indicate
a lack of central symmetry of the halo [13]. Direct
kinematic observations of various halo objects (globular
clusters, satellites, blue horizontal-branch stars,
red giants) can be used to find the radial velocitydispersion
profile which, in turn, indicates the character
of the gravitatingmass distribution for a given halo
model. New possibilities for deriving halo parameters are provided by observations of hyper-velocity
stars [17--19].

Deep surveys of edge-on galaxies indicate a strong
prevalence for stellar halos in the form of oblate
spheroids with axial ratios of  $\sim 0.6$, which have been
attributed to tidal interactions [14, 20]. Cosmological
models taking into account the stellar-halo dynamics
also give triaxial shapes for the dark halo  ($s=0.84$, $q=0.94$) and stellar halo ($s=0.84$,  $q=0.91$)~[21].

Several techniques for determining halo shapes
($a, b, c$) from observations can be identified [22], based
on: (1) expansion of the $HI$ gaseous disk with distance
from the center (flaring of the gaseous layer) ---
other conditions being equal, the thickness of the
gaseous disk decreases with increase in the degree of
flattening of the halo; (2) warping of the gaseous layer;
(3) X-ray isophotes; (4) outer polar rings---a dynamical
model of the polar ring is constructed allowing
for an asymmetric halo and fit to the observational
data; and (5) the parameters of a precessing dust
disk. The observational data for our Galaxy indicate
that the both the stellar and dark-matter halos do not
display central symmetry. Different approaches based
on observations of different objects lead to estimates
of $s \sim 0.2-0.9$ (see, e.g., [23, 24]).

Thus, one of the most important properties of a
halo is the triaxial character of the mass distribution
in it. In turn, information on the character of the darkmatter
distributions in galaxies and in their immediate
vicinities, together with observations of the isotropic
gamma-ray background, can be used to derive constraints
on particle dark-matter models~[25].

Let us consider the dynamics of a rotating gaseous
disk in the non-axisymmetric gravitational field of a
dark halo. Let us assume that themass of the stellar--gaseous disk within the limits of the optical radius $R$
is comparable to the dark mass~[2, 26, 27]. The
influence of the global asymmetry of the potential on
the formation of dynamical structures in the disk is
of the most interest. It seems natural to expect that
the disk responds to this influence by forming a spiral
pattern~[28, 29].

Let us consider the hypothesis that the nonaxisymmetry
(triaxial nature) of the dark halo is
responsible for the formation of the spiral structures
in galactic disks, at least for some galaxies. We will
restrict our consideration here to the gaseous disk. A
separate study will be devoted to taking into account
the stellar disk.

\section{MODEL OF A GASEOUS DISK IN THE
EXTERNAL POTENTIAL}


The character of the spatial mass distribution in
the dark halo determined from observations is uncertain.
Several models for the halo density $\varrho_h$ have been proposed, such as the models of Navarro et al.~[30]
(NFW) and Burkert [31], the exponential model [32],
and the quasi-isothermal halo model [1, 33] (iso).
We will generalize this last model for the case of a
triaxial halo, by writing the following expression for
the potential 
\begin{equation}\label{Eq-Elista-potent-halo}
    \Psi_h(x,y,z) = 4\pi G \varrho_{h0} a^2\cdot \left\{
\ln(\xi) + \frac{{\rm arctg(\xi)}}{\xi} + \frac{1}{2} \ln
\frac{1+\xi^2}{\xi^2}
    \right\} \,,
\end{equation}
where $\Oo\xi=\sqrt{ \frac{x^2}{a_x^2} + \frac{y^2}{a_y^2} +
\frac{z^2}{a_z^2}}$, and $a_x$, $a_y$, and $a_z$ are
characteristic scales along the corresponding axes.
The Poisson equation gives the density distribution 
$$
    \varrho_h(x,y,z) =  \frac{\varrho_{h0}a_x^2}{\xi^2}\left\{\left(
\frac{1}{a_x^2} + \frac{1}{a_y^2} + \frac{1}{a_z^2}
    \right)\left[ 1 - \frac{\arctan(\xi)}{\xi}\right] -   \right.
$$
\begin{equation}\label{eq-Elista-density-halo} \left.
- \frac{1}{\xi^2}\left( \frac{x^2}{a_x^4} + \frac{y^2}{a_y^4} + \frac{z^2}{a_z^4} \right)
 \left[ 2 + \frac{1}{1+\xi^2} - 3\frac{\arctan(\xi)}{\xi} \right] \right\}
 \,.
\end{equation}
In the limiting case of a centrally symmetric halo $a=a_x=a_y=a_z$, we have a quasi-isothermalmodel halo with the density distribution 
\begin{equation}\label{eq-iso-halo}
    \varrho(r) = \frac{\varrho_{h0}}{1+(r/a)^2} \,,
\end{equation}
from (1), (2); this density distribution ensures constancy
of the rotational velocity at large distances
$r\gg a$, where the rotation curve has a plateau. The
halo models differ in the circular velocity $V_c$ at large
distances ($r\gee R_{\max}$, $R_{\max}$ is the boundary of the
stellar disk)~[34]. In the inner region of the galaxy,
we have $V_c^{(iso)}<V_c^{Bur}<V_c^{NWF}$~\cite{Sofue-2009!rotation-curve-Galaxy}. Therefore, other
things being equal, our choice of the model (\ref{Eq-Elista-potent-halo}) should
be considered to yield a lower limit for the influence of
the non-axisymmetric halo.

Below, we restrict our consideration to the model
with $a_x\ne a_y= a_z$. If the halo is flattened in the vertical
direction $(a_z<a_{x,y})$, this effectively increases
the halo mass, since the gravitational force of the
halo in the $z=0$ plane becomes greater, so that the
case $a_z=a_x$ again gives us a lower limit. When
$q=a_x/a_y\ne 1$ in the plane of the disk, we have a
non-axisymmetric halo that can, in general, rotate
with a small angular velocity $\Omega_h$. Observational data
and cosmological models allow values $\varepsilon_h = |q-1|\simeq 0- 0.2$ [8--10].

The character of the flow in the gaseous disk
depends considerably on the values of the indicated model parameters. However, the generation of spiral structure is observed in all cases, and we have a longlived,
quasi-stationary, two-armed wave in the main
region of the disk $(0.5 \lesssim r \lesssim 2)$. Figure 2 shows the
typical long-term evolution of the disk ($t=40$ corresponds
to six rotations of the disk periphery $r\sim 2$).
The two-armed perturbation covers almost the entire
disk, and is trailing except at the center itself, where
more complex structures are observed.

Let us assume that the rotational angular velocity
of the halo $\Omega_h$ is less than the rotational angular
velocity of the outer edge of the disk, $\Omega(R)=V(R)/R$. If typical values of the linear velocity $V(R)$
for galactic disks in the plateau region are $V_{\max}\simeq 100-300$ km/s at distances $R\simeq 10-15$ kpc $(\Omega(R)\sim 10-30$ km s$^{-1}$kpc$^{-1}$), this implies for the angular velocity
of the halo $\Omega_h\lee 5$ km s$^{-1}$kpc$^{-1}$.

Let us consider the complete system of gasdynamical
equations in the form
\begin{equation}\label{Eq-gas-sigma}
  \frac{\partial \rho}{\partial t} + \nabla \cdot (\rho \textbf{u}) = 0 \,,
\end{equation}
\begin{equation}\label{Eq-gas-uvw}
  \frac{\partial \rho \textbf{u}}{\partial t} + \nabla \cdot (\rho \textbf{u}\textbf{u}^{T} + {P} \hat{I} ) = -\rho\nabla\Psi \,,
\end{equation}
\begin{equation}\label{Eq-gas-E}
    \frac{\partial E}{\partial t} + \nabla \cdot ( [E + {P}]\textbf{u} ) = -\rho \textbf{u} \cdot \nabla\Psi \,,
\end{equation}
where $t$ is the time, $\rho$ the gas density, $P$ the pressure,
$\textbf{u} = [u,v,w]^{T}$ the velocity vector, $\hat{I} = \textrm{diag}[1,1,1]$ a
$3\times3$ unit tensor, and $\Psi=\Psi_0+\Psi_g$ the gravitational
potential, composed of an external component (halo
+ stellar disk) $\Psi_0(\textbf{r},t)$ and the potential of the selfgravitating
gas $\Psi_g(\textbf{r},t)$. The total energy per unit
volume of the gas is
\begin{equation}\label{Eq-state-idealgas}
    E = \rho \left(e + \frac{\textbf{u}^2}{2}\right) \,,
\end{equation}
where $e$ is the specific internal energy. The equation
of state of the gas is defined by \mbox{$e = P/(\rho(\gamma - 1))$}, where $\gamma$
is the adiabatic index. The system of equations (\ref{Eq-gas-sigma})--(\ref{Eq-state-idealgas}) must be supplemented with the Poisson equation
$\Delta\Psi_g=4\pi G\varrho$, to take into account the self-gravity.


We used total variation diminishing (TVD) version
of the MUSCL scheme [35, 36] to numerically model
the system of gas-dynamical equations (4)--(6). We
approximated the equations using the finite-volume
method, which ensures that conservation laws are
obeyed at the discrete level [37]. This numerical
scheme is related to Godunov schemes with secondorder
accuracy in time and third-order in space. The
modified HLLC method [38] was used to calculate
the fluxes of physical quantities through cell boundaries;
this provides a means to model shocks and
contact and tangential discontinuities through the
boundaries. The construction of the scheme with
second-order accuracy in time was carried out using a
predictor--corrector scheme compatible with the condition
that the total variation of the numerical solution
not decrease (the TVD condition [36]). A third-order
approximation in space was used in regions of smooth
flow, and was achieved using the MUSCL procedure
for reconstructing simple quantities (the density,
pressure, and velocity vector) at the boundaries of the
computation cells. This simple and computationally
efficient method makes it possible to obtain shocks
spread over three cells. The corresponding parameter
responsible for the artificial compression was used
to improve the resolution of tangential and contact
discontinuities in the interpolation.

The computations were carried out on cylindrical ($r, \varphi, z$) and Cartesian ($x, y, z$) grids with high
resolution. The dimension of the numerical grids
for the two-dimensional (2D) models reached  $5000\times 5000$ for the Cartesian grid and $(N_r \times N_\varphi)=(2400\times 720)$ (for the polar grid. The grid parameters in the three-dimensional (3D) model were $N_r=600, N_\varphi=360, N_z=200$. A special numerical algorithm was
constructed for the through calculation of the gas--vacuum interface, which preserved the positivity of the scheme [39]. In this case, the matter is situated
within the computation domain. A comparative
analysis demonstrated the extremely small influence of the boundary conditions on the evolution of the constructed models for the gaseous disks.

In order to estimate the influence of the numerical
technique used on the disk dynamics, we also carried
out hydrodynamical calculations using Smoothed
Particle Hydrodynamics (the SPH method) [40, 41].
The SPH algorithm is related to fully Lagrangian
algorithms, and so is free of the need to use the
approximation of a spatial grid; this removes a
considerable number of theoretical and algorithmic
difficulties. The scheme is constructed based on
the motion of particles whose evolution in time and
space directly reproduces the conservation of mass,
momentum, and angular momentum. A modification
of the standard predictor--corrector scheme was used
for the time integration of the SPH equations; this
modification ensures second-order accuracy in time.
The calculations were carried out $N=10^5-10^6$ particles,
and demonstrated good agreement with solutions
based on the grid TVD method. The selfgravity
was taken into account using the TREEcode
algorithm.

\begin{figure}
\setcaptionmargin{5mm}
\onelinecaptionsfalse
 \includegraphics[width=0.5\hsize]{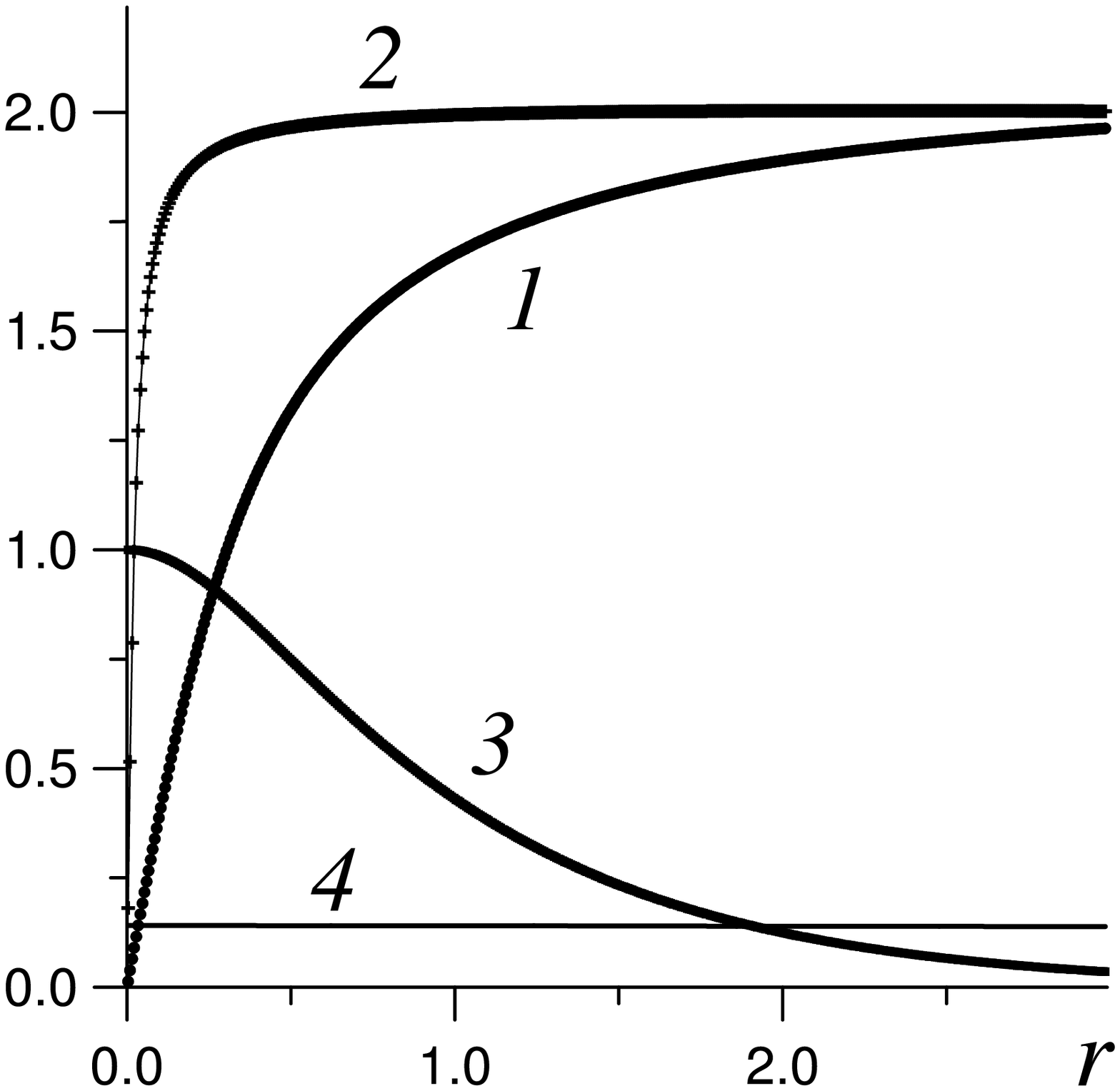}
\captionstyle{normal}
\caption{Typical dependences of the ($1, 2$) rotational velocity
$V (r)$ (without and with a bulge, respectively), (3) surface
density $\sigma(r)/\sigma(0)$, and (4) sound speed $c_s(r)$ on the
radial coordinate for the initial equilibrium state of the
gaseous disk.}
\label{Fig-init-state}
\end{figure}

The initial equilibrium state of the gaseous disk is
defined by the radial and vertical profiles of the rotational
velocity $V$, density $\varrho$, and sound speed $c_s$ in the
absence of radial and vertical motions, $u=w=0$. It
is convenient to characterize the state of the disk using
the Mach number, ${\cal M} = {V}/{c_s}$. Since ${\cal M}$ depends
on the coordinates, it is convenient to introduce the
quantity ${\cal M}_0=V_{\max}/c_s(r=0)$. Figure ~\ref{Fig-init-state} presents
typical radial profiles of the disk parameters at the
initial time for the assumed dimensionless quantities. 
The rotation curves are typical for galactic-disk components,
and are characterized by a small section with
a quasi-rigid-body rise at the disk center and a more
extended section with roughly constant velocity (the
so-called plateau in the rotation curve).

The numerical model assumes $G=1$, and the
galaxy mass within the limits of the radius $r\le 1$ is
$M_{gal}=4$. The optical radius of the galaxy (the outer
edge of the stellar disk) is $R=2$ in dimensionless
units. In this case, a typical value of the dimensionless
rotational velocity of the disk is $V_{\max}\simeq 2$, and the
rotational period of the periphery of the gaseous disk
($r\sim 2-3$) is $T\simeq 6-10$. The gaseous disk extends
beyond $R$.

We considered various initial radial profiles of the
gas surface density $\sigma$:
\begin{enumerate}
    \item[(1)] $\sigma\propto\exp(-r/L)$,
    \item[(2)] $\sigma\propto(1+(r/L_1)^2+(r/L_2)^6)^{-1}$,
    \item[(3)] $\sigma\propto(1+(r/L_3)^2)^{-5/2}$,
    \item[(4)] $\sigma\propto\cosh^{-2}(r/L_4)$.
\end{enumerate}
The calculation domains in the various models are
situated within $R_{\max}=5-15$, where $\sigma(R_{\max})/\sigma(0)= 10^{-5}-10^{-10}$, which ensures an absence of influence from boundaries.

At the initial time $t=0$, the axisymmetric equilibrium
disk is in the axisymmetric potential $\Psi(r,t=0)$ with $\varepsilon_h=0$ ($a_x=a_y$). Then, the value of the scale $a_x$ was increased linearly to $\varepsilon_h=0.001\div 0.2$ over a time $\tau_h$ for the various models. After the nonaxisymmetry was "switched on"  at times $t>\tau_h$, the
halo potential remained constant.

In the following section, we describe the main
results of our modeling of the disks, focusing on the
possible generation of various structures by the triaxial 
halo, considering the morphology and kinematics
of these structures based on the results of over 80
numerical experiments.

\section{GASEOUS DISK IN A NON-AXISYMMETRIC HALO}

%

The main free parameters of the constructed model
and their typical dimensionless values are:

\noindent (1) the degree of non-axisymmetry of the halo $\varepsilon_h=0\div 0.2$ ($q=a_y/a_x=1-\varepsilon_h$).

\noindent (2) the width of the quasi-rigid-body section of
the rotational velocity in the central region of the disk
$d=0.015-0.5$;

\noindent (3) typical scales for the radial distribution of the
surface density $L=0.5-1.5$; 

\noindent (4) the sound speed of the gas or the disk temperature,
characterized by the effective Mach number
${\cal M}_0=5-30$;
 
\noindent (5) the relative gas mass $\mu_g=M_g/M_0$ ($M_0$ is the
galaxy mass within the radius $r\leq R=2$); 

\noindent (6) the rotational angular velocity of the halo $\Omega_h=0-0.2$, corresponding to the corotation radius at the
periphery of the gaseous disk; 

\noindent (7) the transition time from the axisymmetric to
the non-axisymmetric halo $\tau_h=0-30$.

Let us note some typical features of the morphology
and kinematics of the disk indicated by our
numerical simulations:

\begin{enumerate}
    \item In all cases, the presence of a non-axisymmetric
halo leads to the formation of spiral structures in the
gaseous disk (see the distribution of the logarithm
of the surface density in Figs. 3a--3d). Under typical
conditions, perturbations grow to the strongly
nonlinear stage over one to three disk revolutions,
forming a system of shocks (Fig. 3e). The time when
the quasi-stationary spiral structure forms increases
with decreasing $\varepsilon_h$. With decreasing $\varepsilon_h$, the spirals
become more tightly wound and have smaller amplitudes,
other conditions being equal (Fig. 3e). A
massive, non-axisymmetric halo can generate nonlinear
waves even when $\varepsilon_h < 0.01$, but the rise time
in this case exceeds eight rotation periods of the disk
periphery. The formation of the spiral structure in the
essentially nonlinear stage depends only weakly on
the time when the non-axisymmetric part of the potential
is switched on, $\tau_h$, which was varied from $\tau_h=0$ (instantaneous) to an adiabatically slow ramping up
during several rotations of the disk periphery (about
1--1.5 billion years for a typical \textit{S} galaxy).
\item
The analysis of the gas-flow structure indicates
a complex character that includes a system of shocks
and of regions of strong shear flows (Fig. 4). The
nonlinear (shock) waves have large-scale structure,
which encompasses the disk right to small densities
$\varrho/\varrho_{\max}\sim 10^{-3}$. The shape of the spiral depends
substantially on the model parameters, but it is a
large-scale formation, formed by rotating through the
angle of $\sim 2\pi$, or, in some cases, by more than $4\pi$.
The amplitude of the spiral waves depends on the
degree of non-axisymmetry of the halo $\varepsilon_h$, but the
formation of the shocks occurs for all of the parameter
values considered. A zone of shear flow is formed in
the region of the spiral density wave near the shock
front (Fig. 4a). The region of strong variation of the
tangential velocity component is large compared to
the shock front, whose width $\ell$ is determined by the
properties of the numerical model (Fig. 4b). The
quantities $\sigma$ and $u$ exhibit a jump at the shock front
(within $\ell$), in contrast to $v$, whose region of variation
is substantially larger than $\ell$. The width of the shock
front $\ell$ decreases with increasing resolution of the
numerical scheme (increasing numbers of cells $N_r$, $N_\varphi$), so that the shock front is described by three to
four numerical cells; this was achieved by using a
numerical scheme with second-order of accuracy in
time and third-order accuracy in the spatial coordinates.
The width of the shear flow region is finite and
does not depend on the choice of parameters for the
numerical grid.


%
\item The geometry of the spiral patterns depends
on the gas-rotation curve $V(r)$, the character of the
density distribution in the dark halo $\varrho_h(r,\varphi)$, and the
radial profiles of the disk surface density $\sigma(r)$and the
sound speed $c_s(r)$. Leading spirals, which always become trailing spirals at the periphery and produce
complex $\Theta$ structures (embedded $\Theta$ structures are
possible) at the center, can be formed in the central
part of the disk (Fig. 5). In all the figures presented
here, the gaseous disk rotates in the counterclockwise
direction. Thus, trailing spirals untwist in the clockwise
direction, and leading spirals in the counterclockwise
direction. The presence of leading spirals
in the very central zone of the disk is clearly visible in
Figs. 3d, 5c and Figs. 9a, 9b, 10a.
\item The generation of pronounced $\Theta$ structures
formed by leading spirals in the central region occurs
in gaseous disk models without allowance for
the perturbation of the stellar-subsystem?s density.
Since $\Theta$ structures are not formed in models with
collisionless stellar disks, the central $\Theta$ structures will
be weaker in self-consistent stellar-gaseous systems,
and the question of the dynamics of the central region
requires a separate study. Note that complex gaseous
structures manifest as, e.g., dust streaks inside the
bar are clearly visible in images of SB galaxies (e.g.,
NGC~1672, NGC~1300, NGC~1365, NGC~3627,
etc.). In addition to gaseous?dusty spirals emerging
from the center, on the whole, along the major axis,
structures tracing the shape of the bar itself (e.g.,
in galaxies NGC~1097, NGC~1288, NGC~2903)
or outer spiral arms that do not start from the bar
ends but encompass part of the leading edge of the
bar (NGC~4548,VCC~1615) are observed. The socalled
three-kpc arms in our Galaxy may serve as an
example.

%
\item 
As in the case of decreasing $\varepsilon_h$, decreasing the
zone of the quasi-rigid-body rotation in the center
of the gaseous disk promotes the formation of more
tightly wound spirals. Simultaneously, the conditions
for the generation of $\Theta$ structures in the center become
less favorable. Trailing spirals now arise in the
center itself and encompass the entire disk (Fig.~\ref{fig-10-spiral-Theta}).

\item In cooler disks (with higher Mach numbers),
more tightly wound and thinner spirals are formed in
the nonlinear stage (Fig.~\ref{fig-10-spiral-Mah}).

\item  The typical spiral pitch angles are $i\simeq 5^\circ\div 20^\circ$.
More open patterns do not formin themodels considered.
Figure 7 shows an example of an approximation
of the obtained structure with sections of logarithmic
spirals $r=r_a \exp(\chi\varphi)$  with pitch angles$i=\pi/2-\arccos(\chi/\sqrt{1+\chi^2})$. For the case presented,$i$ increases from $10^\circ$ in the central zone to $14^\circ$ at the
periphery. This small increase in $i$ with radius is
typical for the models constructed.

\item Let us now consider the rotational angular velocity
of the spiral pattern $\Omega_p$. Figure 8 shows the time
dependences of $\Omega_p$ for the two-armed spiral at various
radii~$r$. An initial stage in the pattern formation can be
distinguished ($t\lee 10$; these processes proceed faster
nearer the center than at the periphery), after which a quasi-periodic rotation regime for the non-linear perturbations develops. We emphasize that the spiral pattern rotates even in the case of a stationary halo
$\Omega_h=0$. However, the rotational angular velocity is then low, and the corresponding corotation radius is
beyond the disk. Computations with a slowly rotating
halo, $\Omega_h\lee 0.2$, show that the geometry and
kinematics of the spiral patterns depend weakly on $\Omega_h$. A small increase of the angular velocity of the wave in the disk is observed with increasing $\Omega_h$ \footnote{Note that the question of the location of the corotation radius is far from resolved, and this is one of the key questions in
spiral structure theory.}. Objects whose spiral structure rotates fairly slowly
and whose corotation radius is at the periphery of, or
even outside, the outer limits of the spiral structure
(e.g., in NGC 3359 [42]) are of interest for the spiralstructure
formation mechanism discussed here.

\item There is qualitative agreement between the results
of the 2D and 3D gas-dynamical computations
(Figs. 9, 10). In the initial stages of formation of the
spiral pattern (over several rotations), the formation of
strongly nonlinear wave fronts propagating in the vertical direction at an angle to the plane of the gaseous
disk is observed. The geometry of the trailing spiral
pattern in the main part of the disk differs slightly in
the 2D and 3D models (Figs. 9, 10). Discrepancies
are observed only in the central region, where the
formation of complex, small-scale structures is possible.
In this case, the presence of a vertical dimension
can substantially change the gas dynamics, since the
sound speed in the center is known not to be small
compared to the rotational velocity of the system, and
vertical motions are of considerable importance \cite{Gorjkavyj-Fridman-1994!Book}.

\item On the whole, taking self-gravity into account
increases the efficiency of the formation of the spiral
pattern by the non-axisymmetric halo. We have considered
only gravitationally stable disks. A change of the wave geometry is observed, and the wave amplitude increases (Fig. 11).
\end{enumerate}

In conclusion, let us summarize the main differences
of the spiral-pattern modeling in a nonaxisymmetric
halo and in the presence of a central bar:

(1) the bar is a rapidly rotating structure relative to
the spiral pattern;

(2) the dimensions of the central stellar bar are
small compared to those of the galactic disk, and the
effect of the bar is dominant at the center and rapidly
decreases with radius; the triaxial halo influences the
entire disk, and influences the disk center only to a
smaller degree;

(3) the degree of non-axisymmetry of the halo is
substantially less than of that the bar, so that the
geometry of the spiral waves differs considerably from
the disk in the nonaxisymmetric halo, even inside the
bar, where the gas, on average, rotates faster than the
bar.

The results of our numerical modeling show that
a triaxial (non-axisymmetric in the plane of the disk)
halo can lead to the generation of a spiral pattern in
the galactic disk, whose properties depend strongly
on the parameters of the disk and halo.

The question of whether non-axisymmetric halos
are responsible for the real spiral patterns observed in
galaxies is, of course, far from being resolved. The
real picture can be made considerably more complex
by the formation of transient spirals, a central stellar
bar, or global modes [44] due to the development of
gravitational instability. In any case, the models considered
here can be used to place some constraints on
the shapes of dark halos in the plane of their galactic
disks. If the properties of the observed spiral pattern in
a galaxy are incompatible with our modeling results,
then this indicates a high degree of axisymmetry for
the dark halo in the plane of the disk.
 
\section*{ACKNOWLEDGMENTS}
The authors thank A.V. Zasov and V.I. Korchagin
for numerous discussions. The numerical
computations were performed on the SKIF
MSU "Chebyshov" supercomputer with the assistance
of A.V. Zasov and N.V. Tyurina. This work
was supported by the Russian Foundation for Basic
Research (projects 07-02-01204, 09-02-97021)
and the Federal Targeted Program "Scientific and
Scientific?Pedagogical Staff of Innovative Russia"
[grant 2009NK-21(7)].

\centerline{REFERENCES}
%


\newpage

\newpage
%
\begin{figure}
\setcaptionmargin{5mm}
\onelinecaptionsfalse
 \begin{tabular}{cccc}
\includegraphics[width=0.25\hsize]{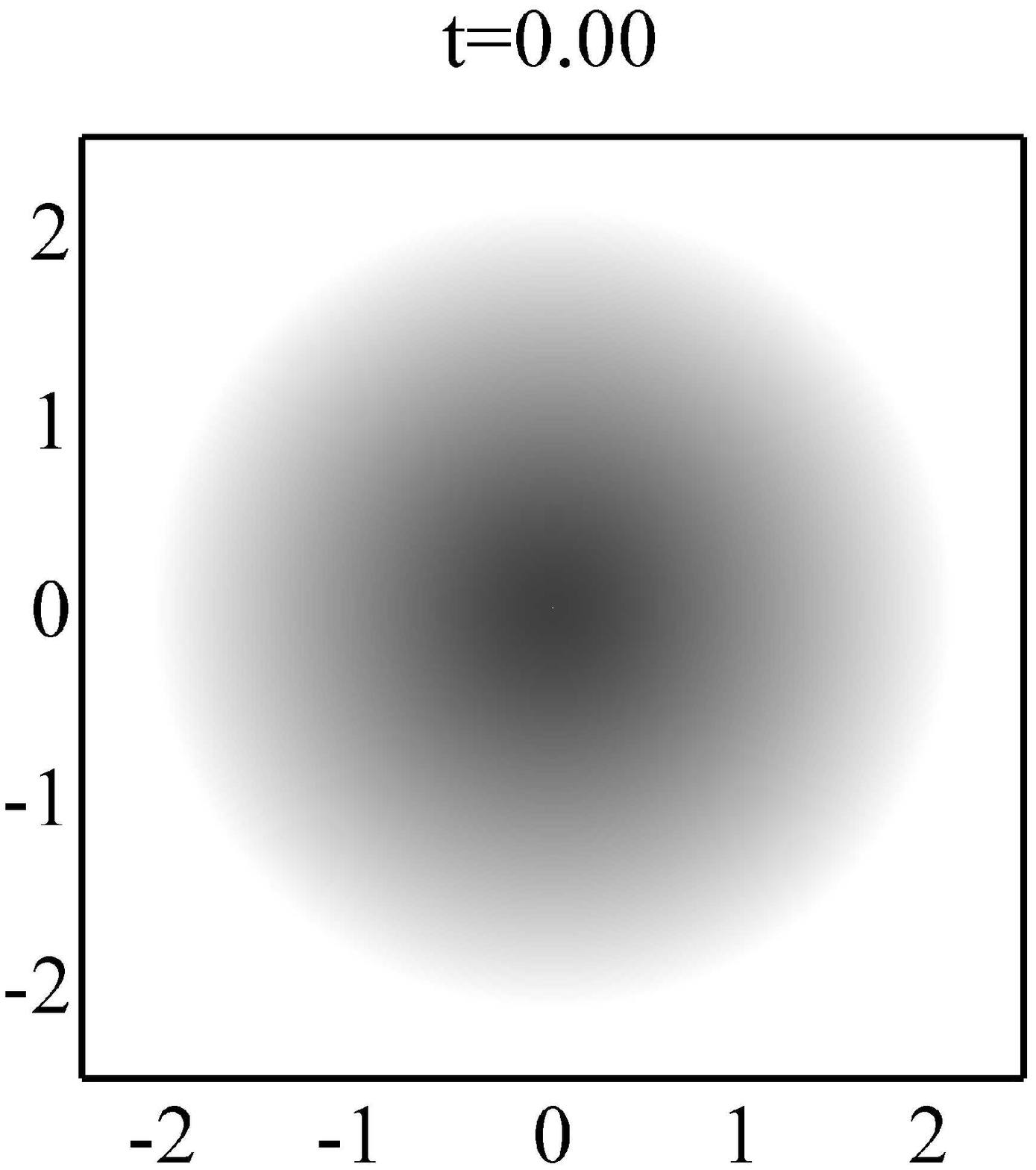} & \includegraphics[width=0.25\hsize]{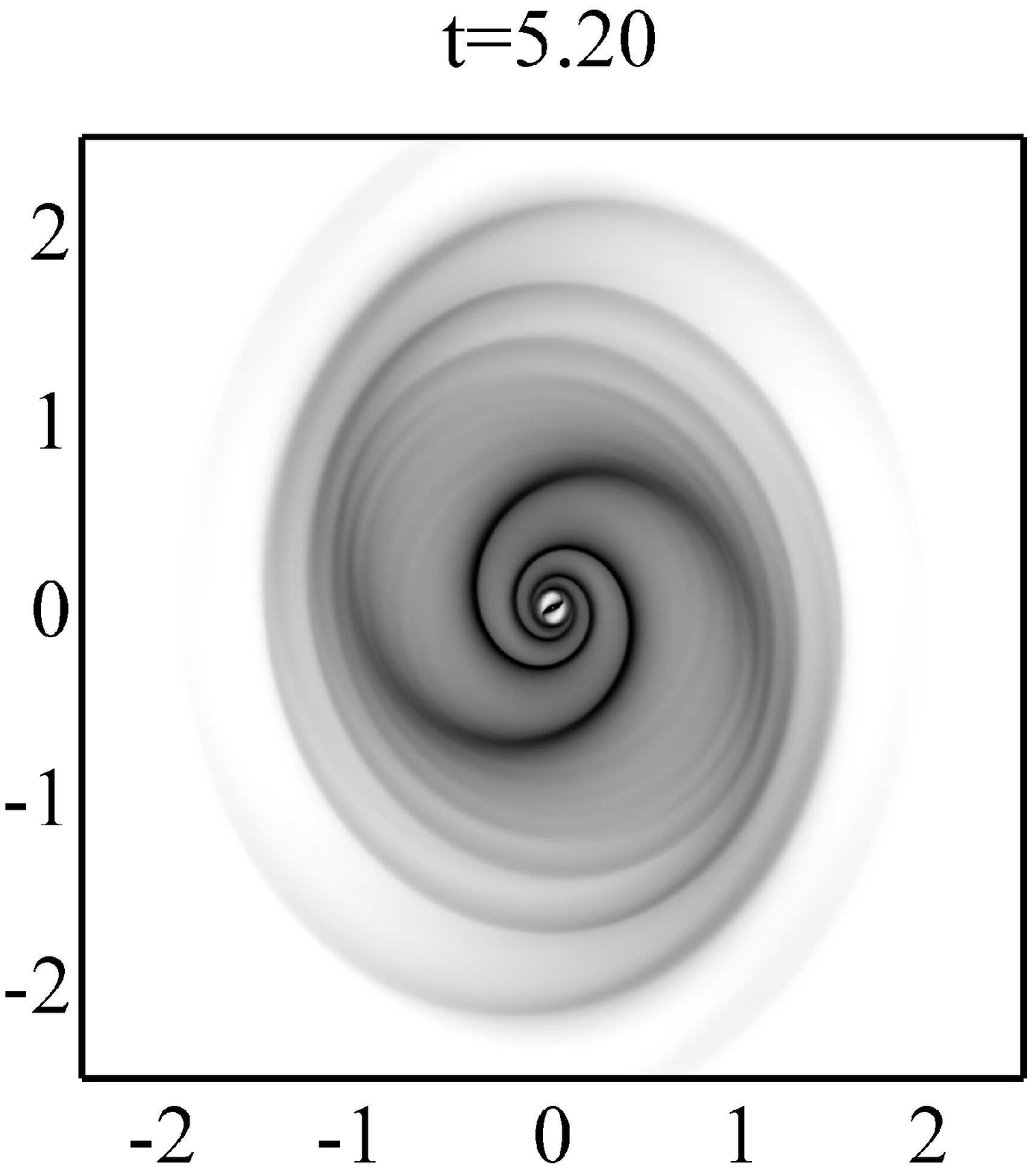} & \includegraphics[width=0.25\hsize]{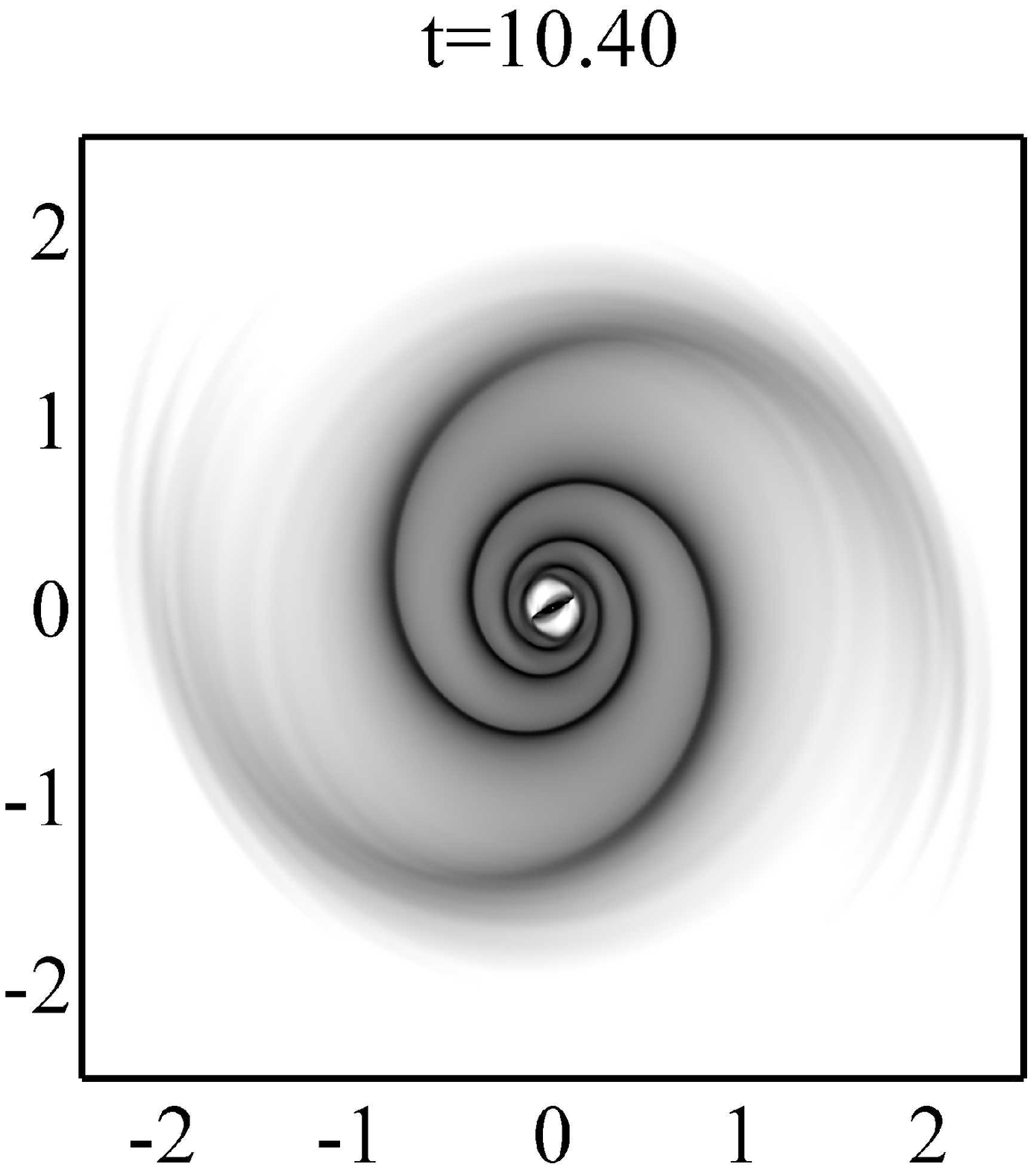} & \includegraphics[width=0.25\hsize]{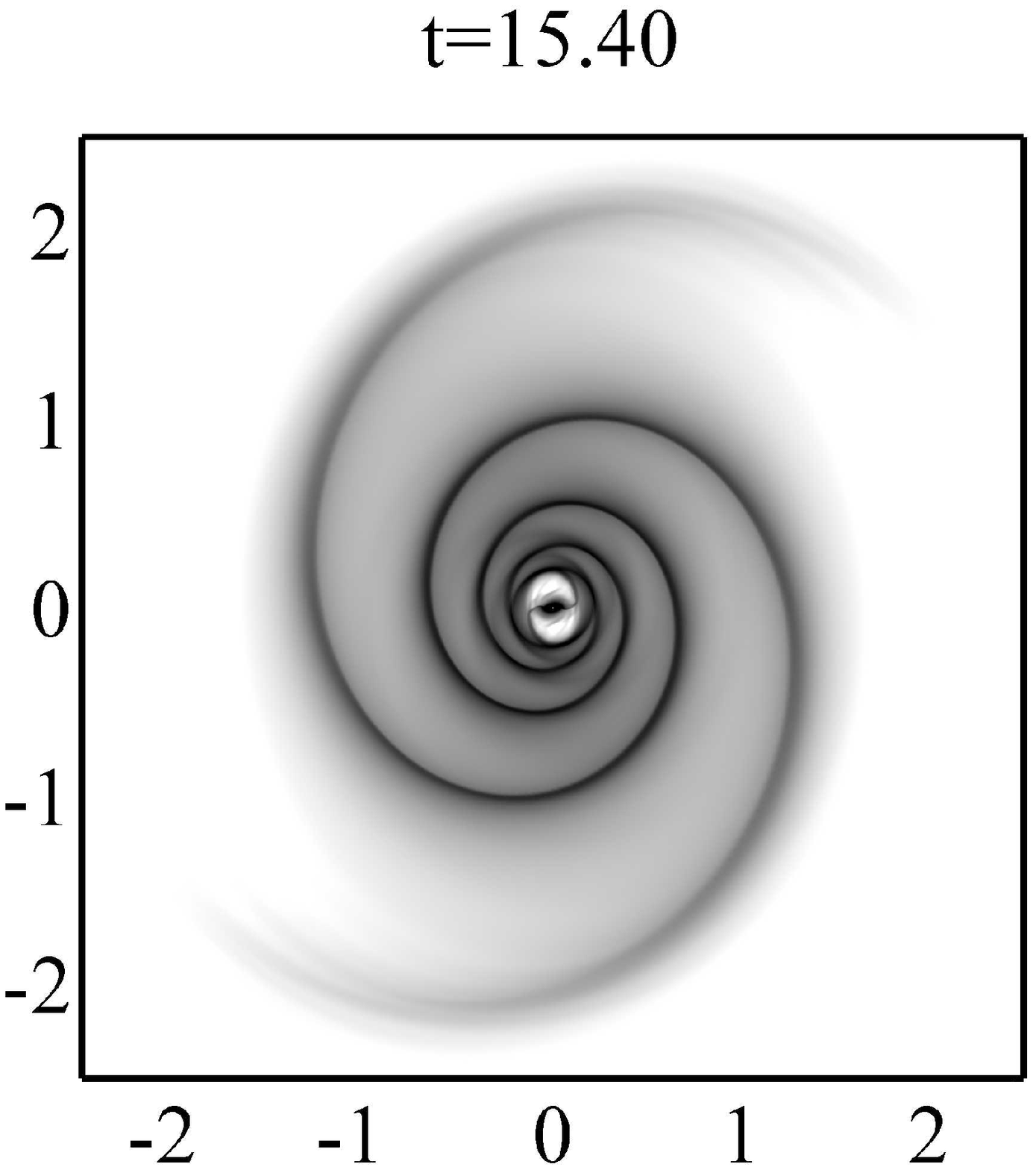} \\
\includegraphics[width=0.25\hsize]{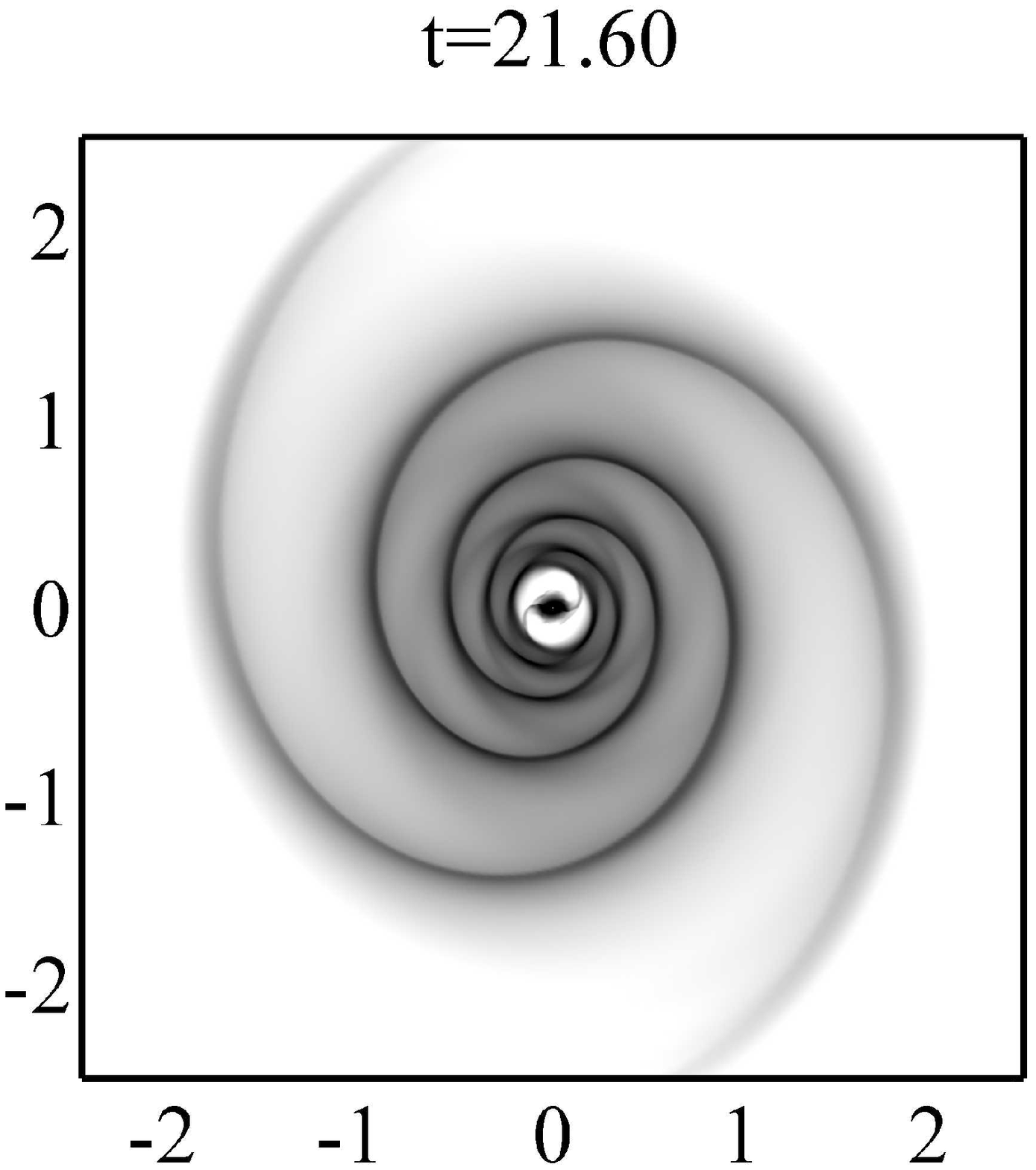} & \includegraphics[width=0.25\hsize]{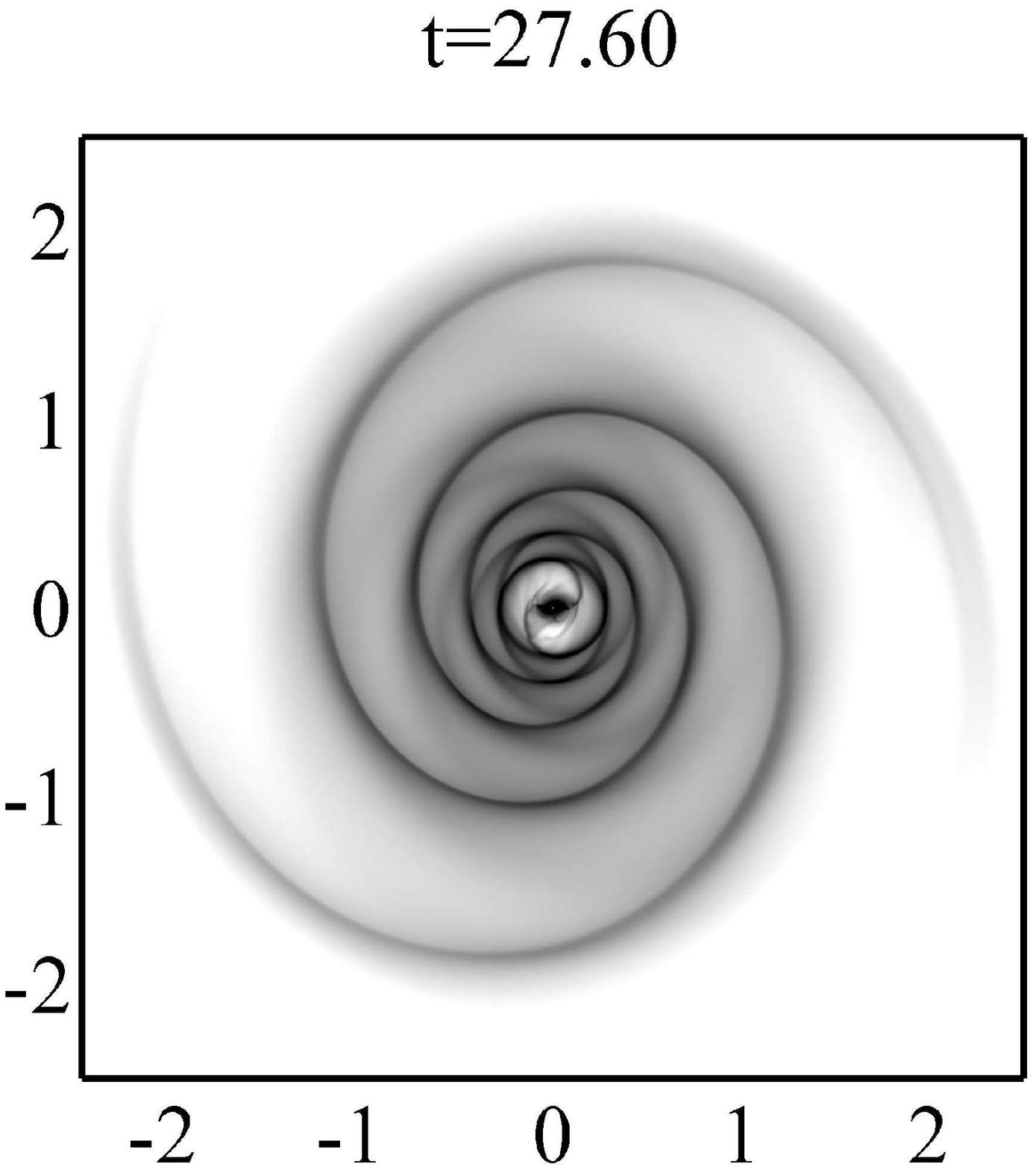} & \includegraphics[width=0.25\hsize]{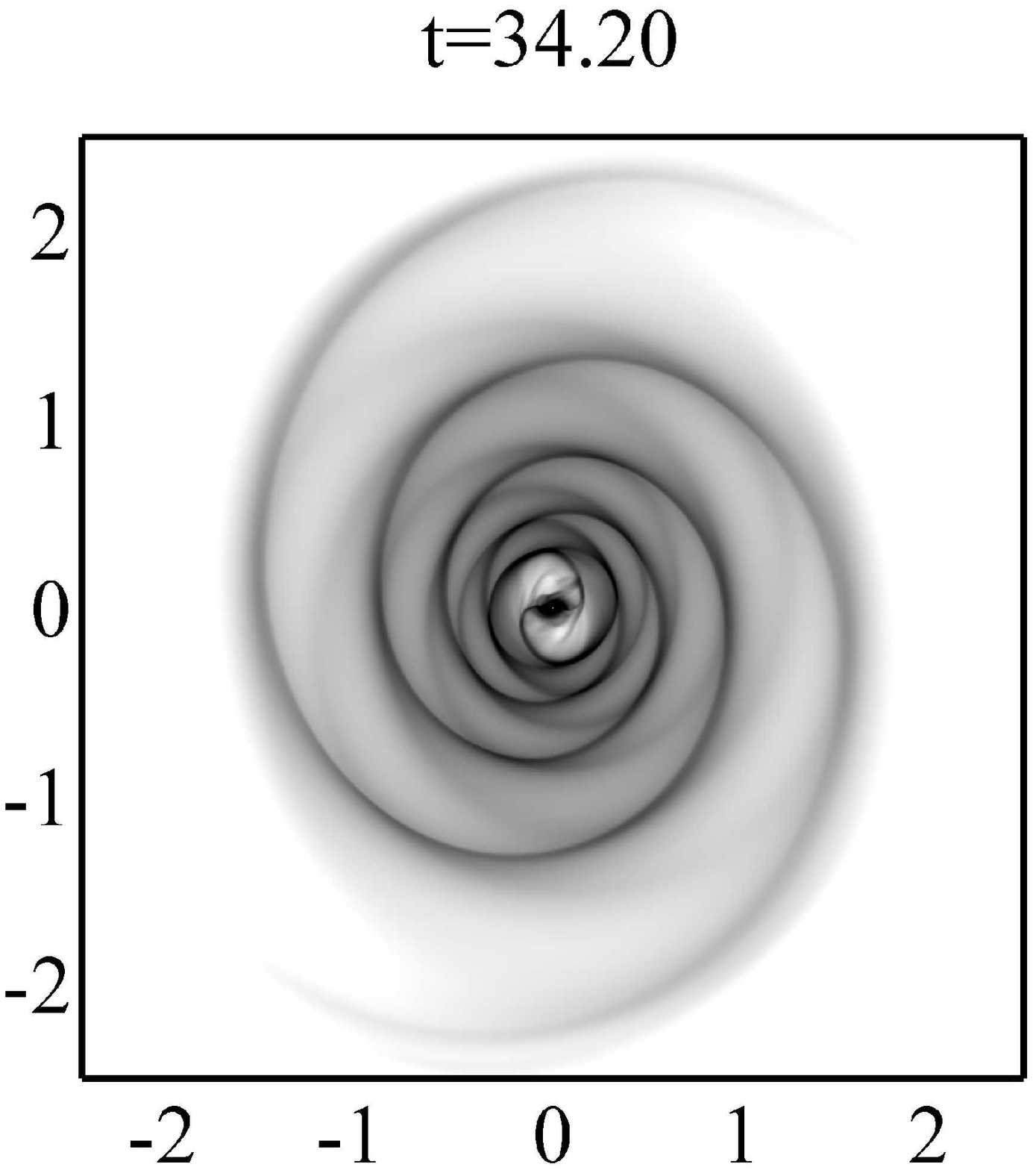} & \includegraphics[width=0.25\hsize]{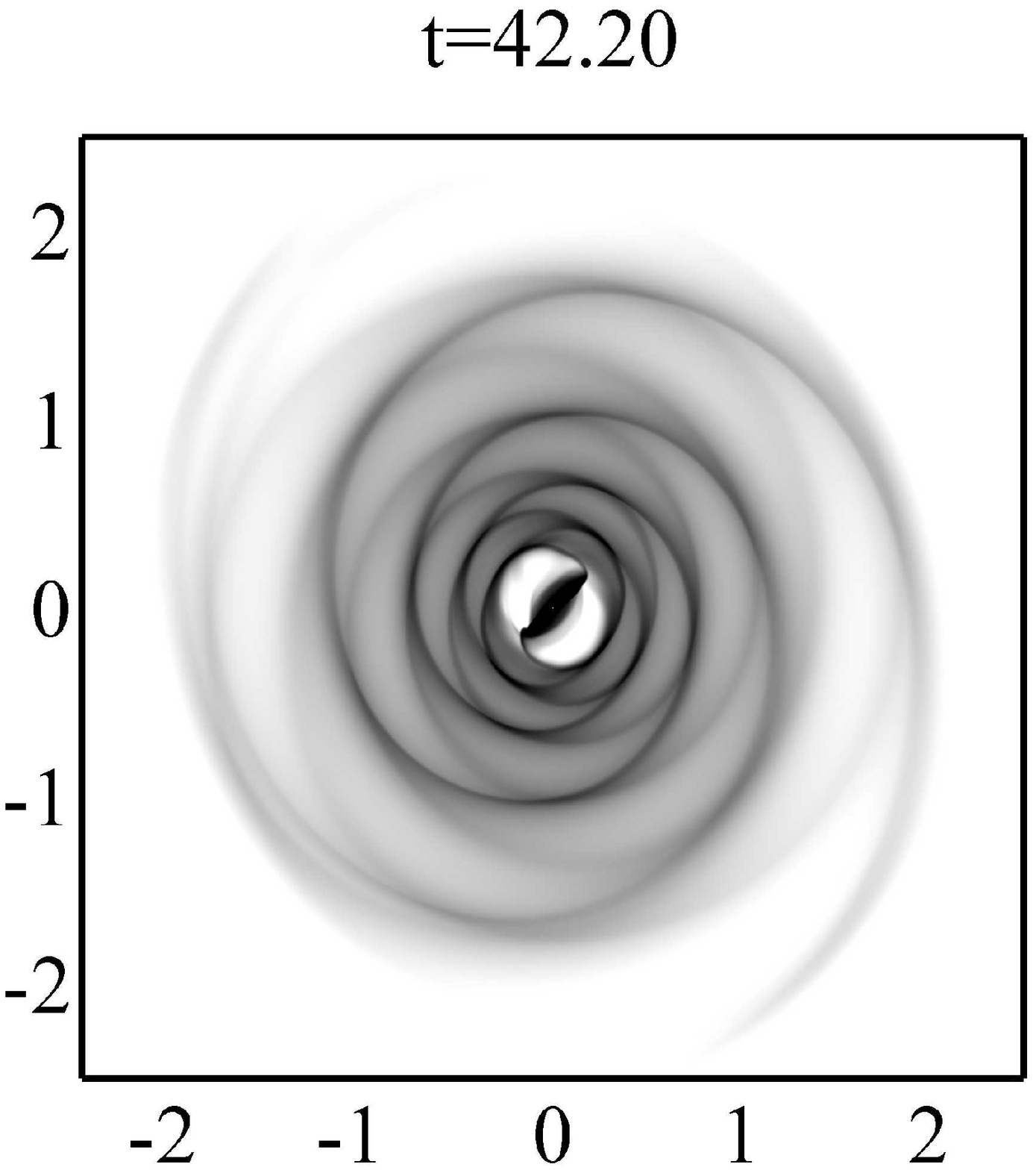}
\end{tabular}
\captionstyle{normal}
\caption{Distributions of the logarithm of the surface density $\sigma$ in the plane of the disk at various times, for the model with $d=0.02$, $\varepsilon_h=0.1$, ${\cal M}_0=10$, and $L_3=0.7$.}
\label{Fig-evolut-disk}
\end{figure}

\newpage
%
\begin{figure}
\setcaptionmargin{5mm}
\onelinecaptionsfalse
\begin{tabular}{cc}
\includegraphics[width=0.4\hsize]{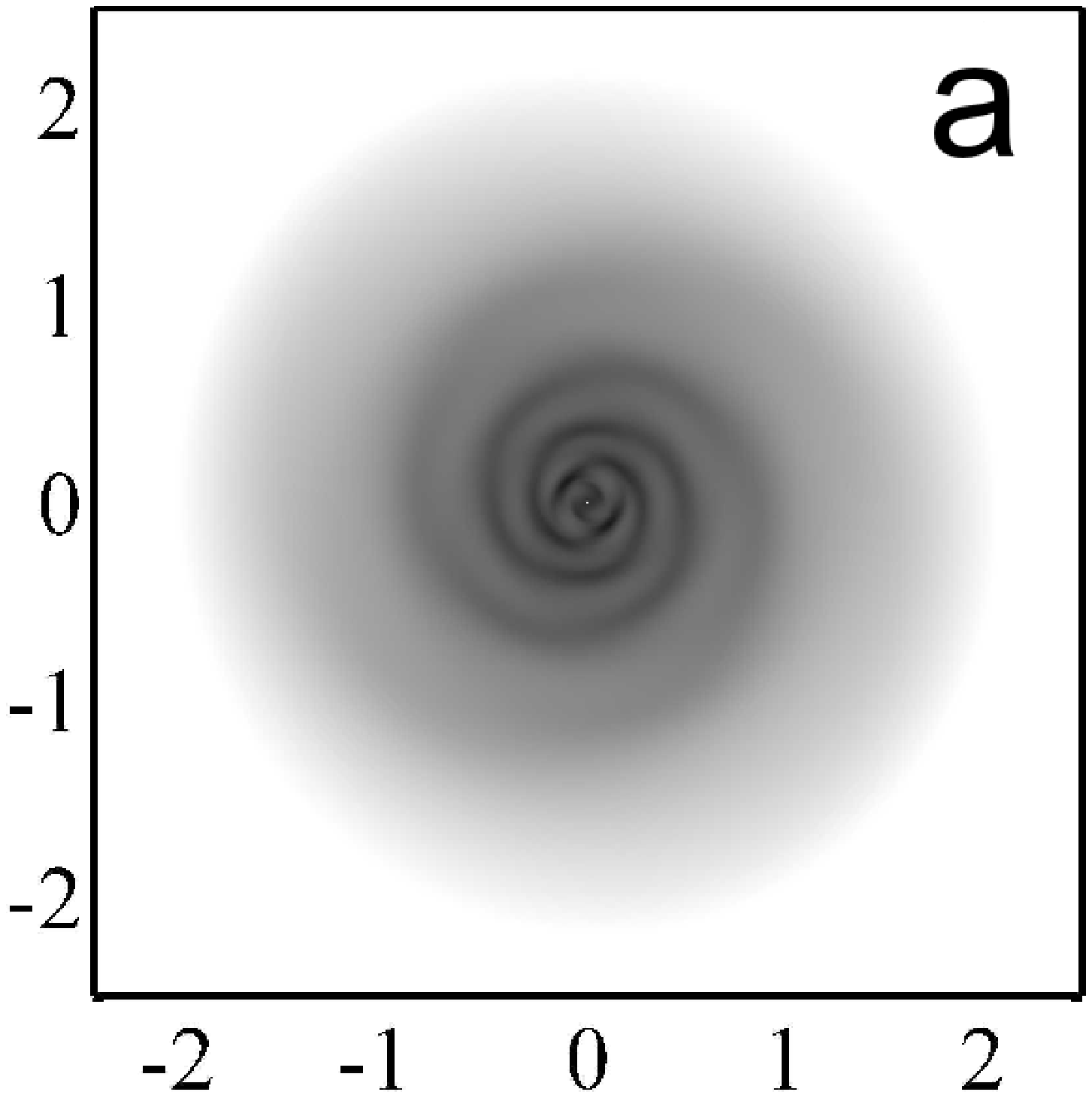} &
\includegraphics[width=0.4\hsize]{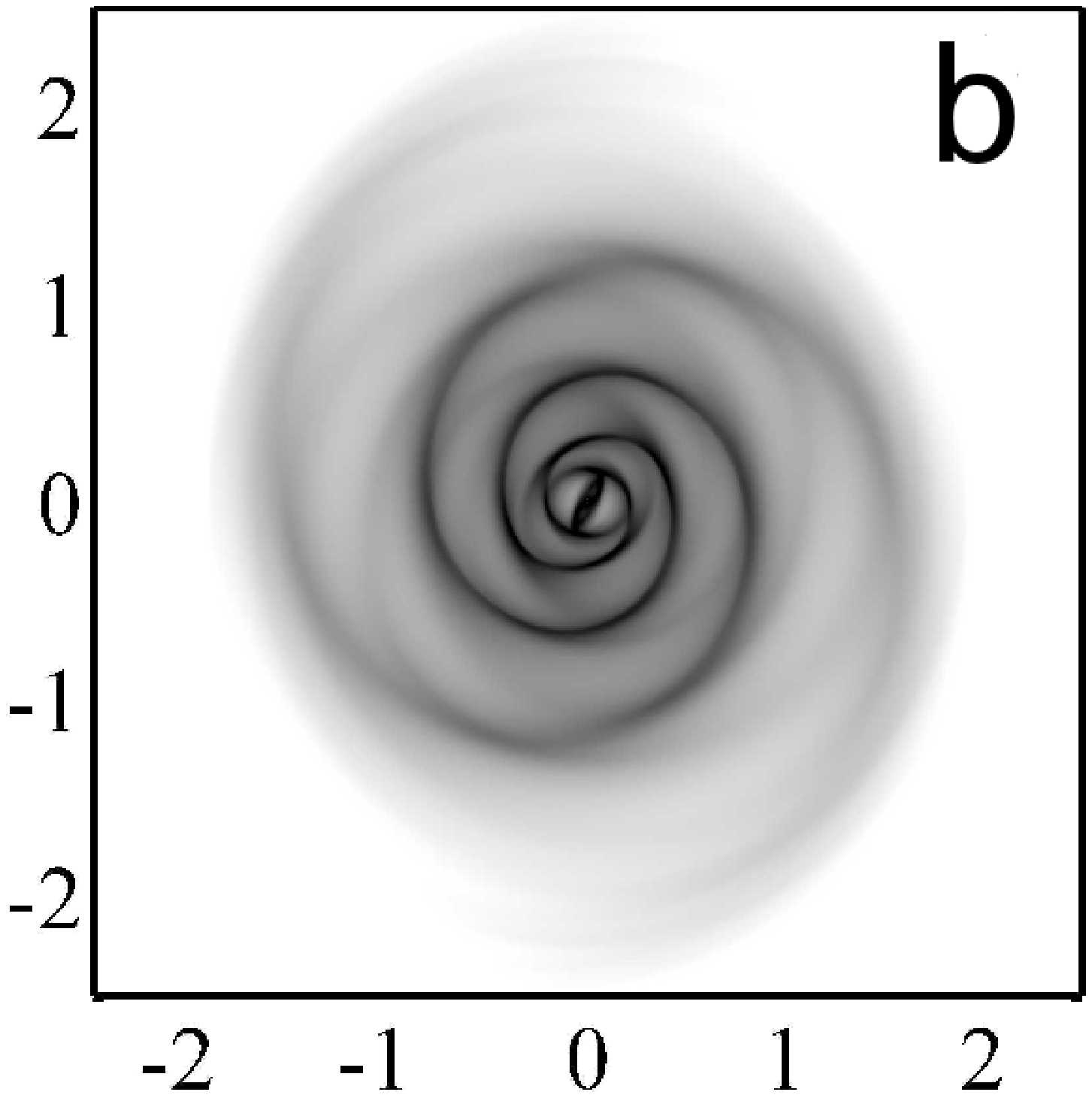} \\
\includegraphics[width=0.4\hsize]{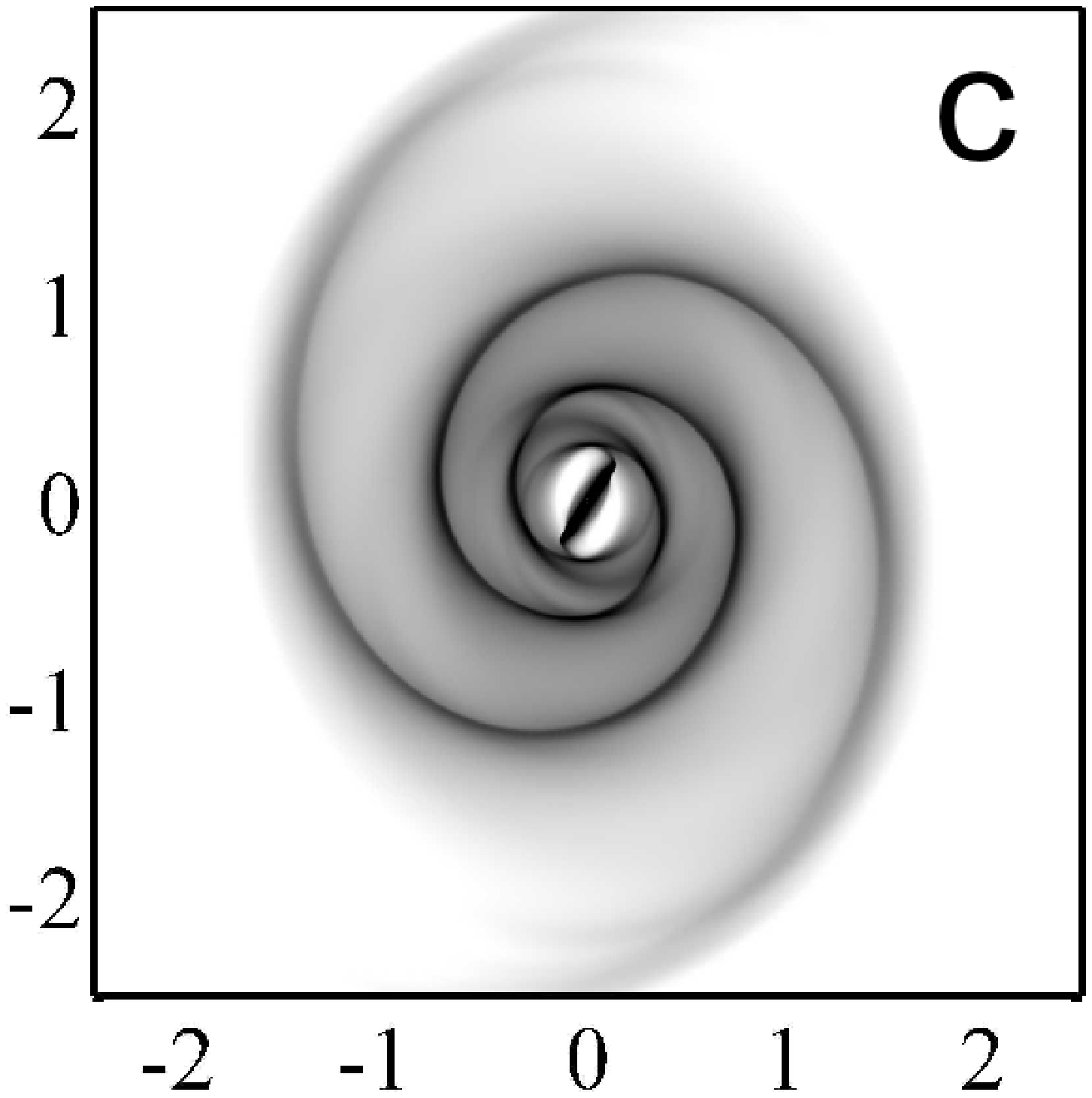} &
\includegraphics[width=0.4\hsize]{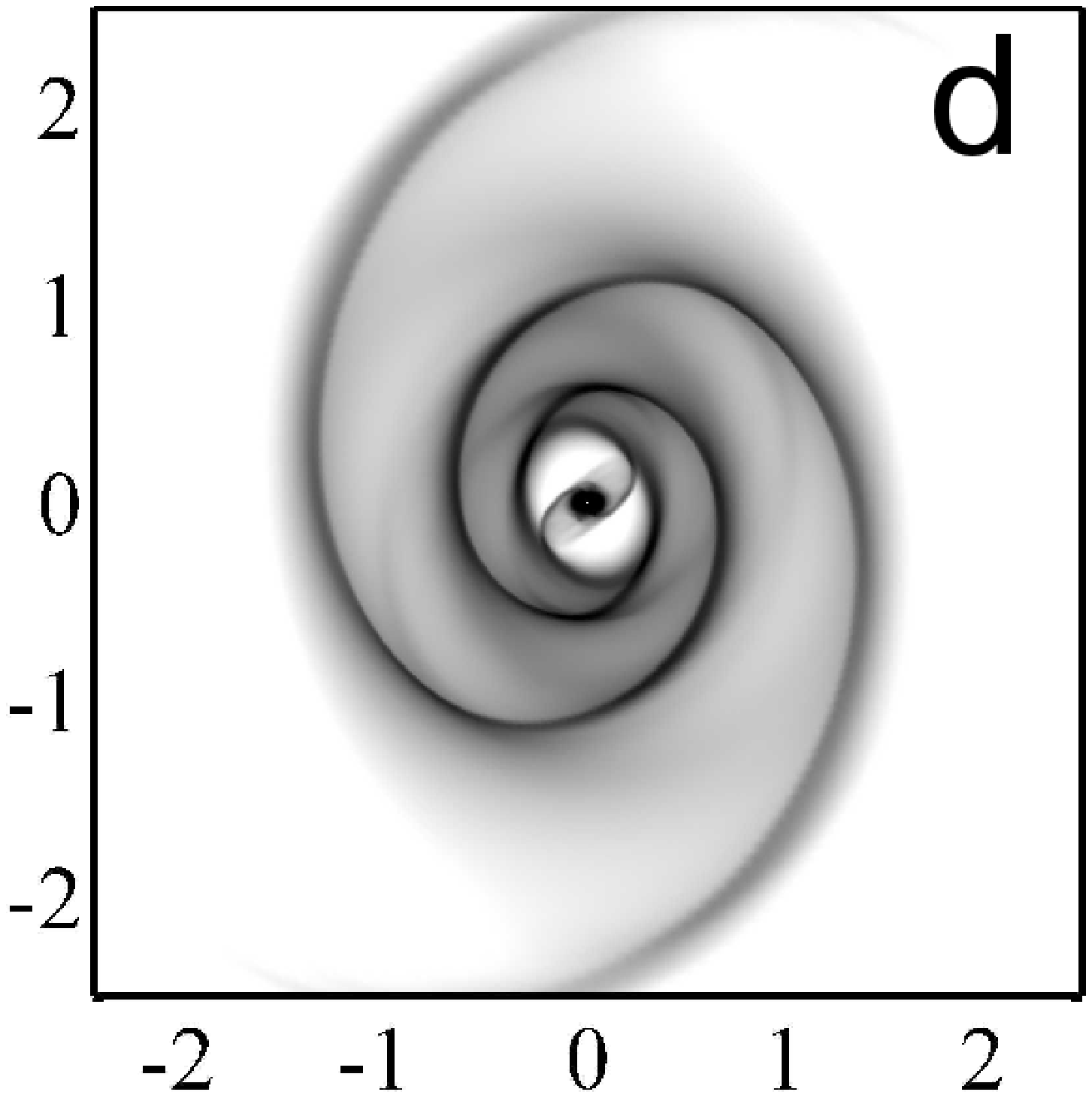}
\end{tabular}
\includegraphics[width=0.45\hsize]{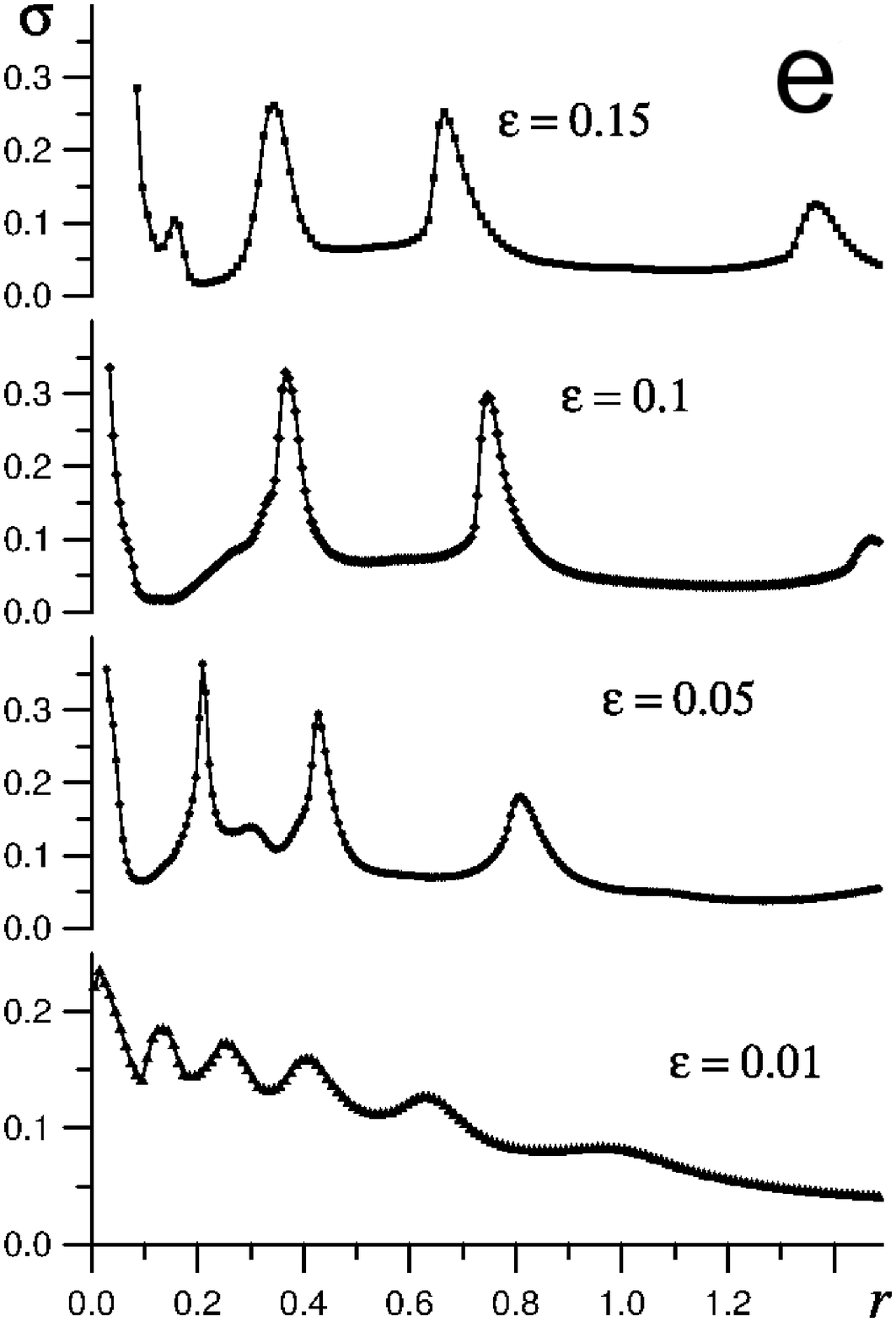}
\captionstyle{normal}
\caption {Distributions of the logarithm of the surface density in the plane of
the disk for models with ${\cal M}_0=10$, $d=0.1$, and $\varepsilon_h=\text{ (a)
}0.01,\text{ (b) } 0.05, \text{ (c) } 0.1, \text{ and } \text{ (d) } 0.15$, all
at the same time. (e) Radial profiles of35 the surface density for the fixed
azimuth angle $\varphi=0$ for the indicated $\varepsilon_h$
values.}\label{fig-10-spiral-eps}
\end{figure}
\newpage
%
\begin{figure}
\setcaptionmargin{5mm}
\onelinecaptionsfalse
            \includegraphics[width=0.6\hsize]{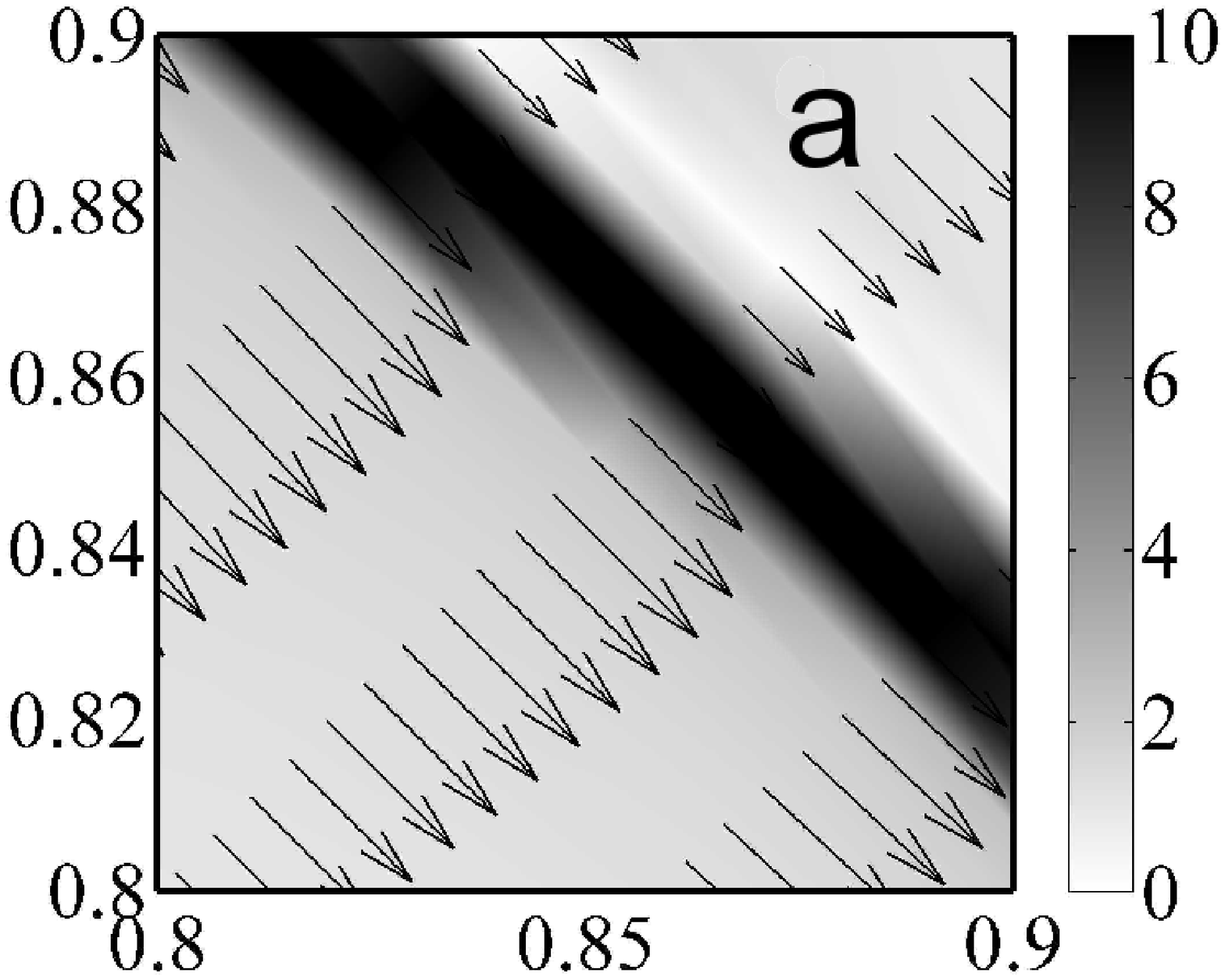}

            \includegraphics[width=0.5\hsize]{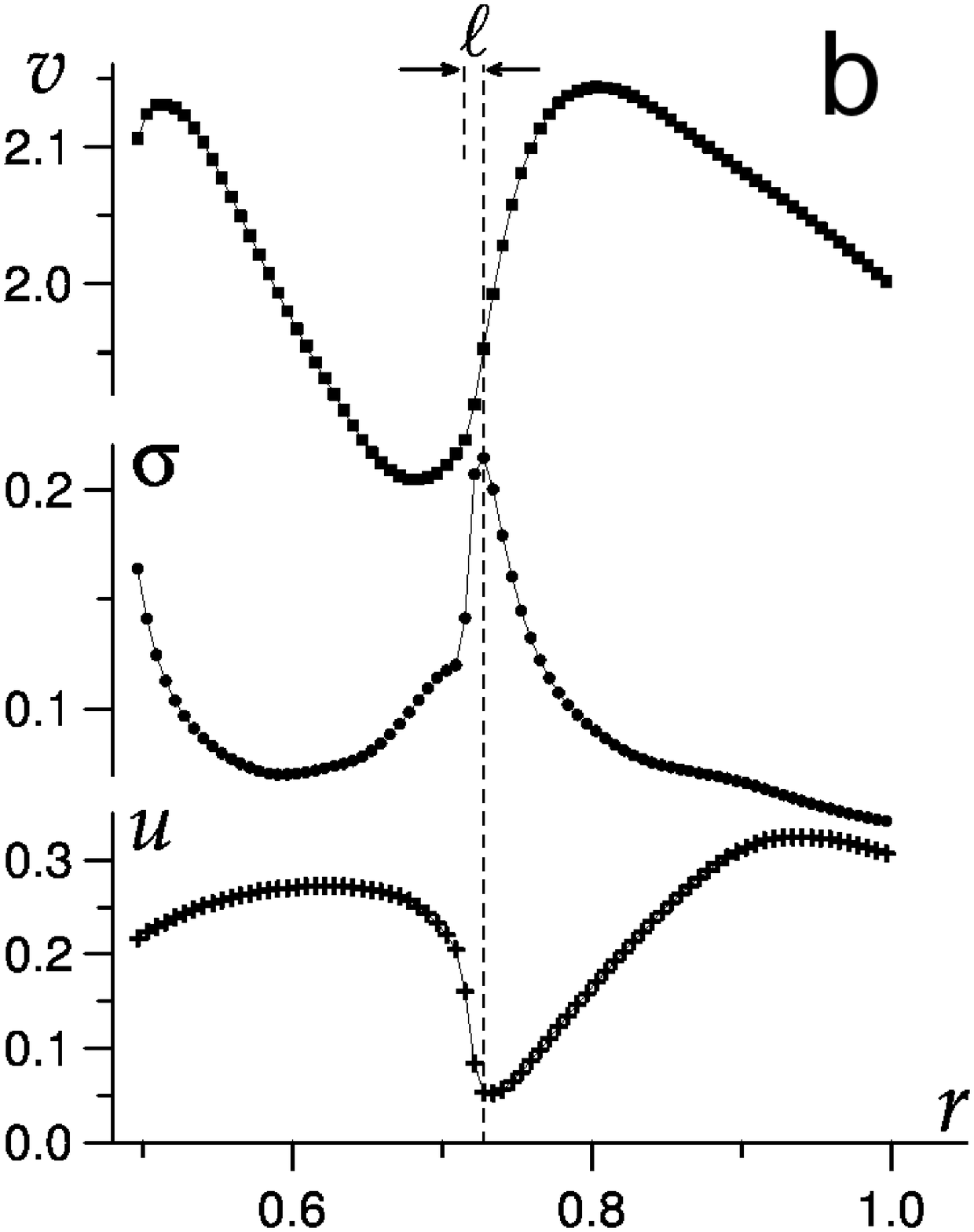}
\captionstyle{normal}
\caption {Model with $\varepsilon_h=0.1$, $d=0.02$, $L_3=0.7$, ${\cal M}_0=10$, and $\Omega_h=0$. (a) Distribution of $\lg((\textrm{div}\,{\textbf{u}})^2)$ (in shades of gray), together with the velocity field ${\textbf{u}}$. (b) Radial profiles of the radial velocity $u$, azimuthal component $v$, and surface density $\sigma$ near
the spiral density wave. The function values at the nodes of the numerical grid are marked near the curves.}\label{fig-10-spiral-gas-div}
\end{figure}
\newpage
%
\begin{figure}
\setcaptionmargin{5mm}
\onelinecaptionsfalse
\includegraphics[width=0.45\hsize]{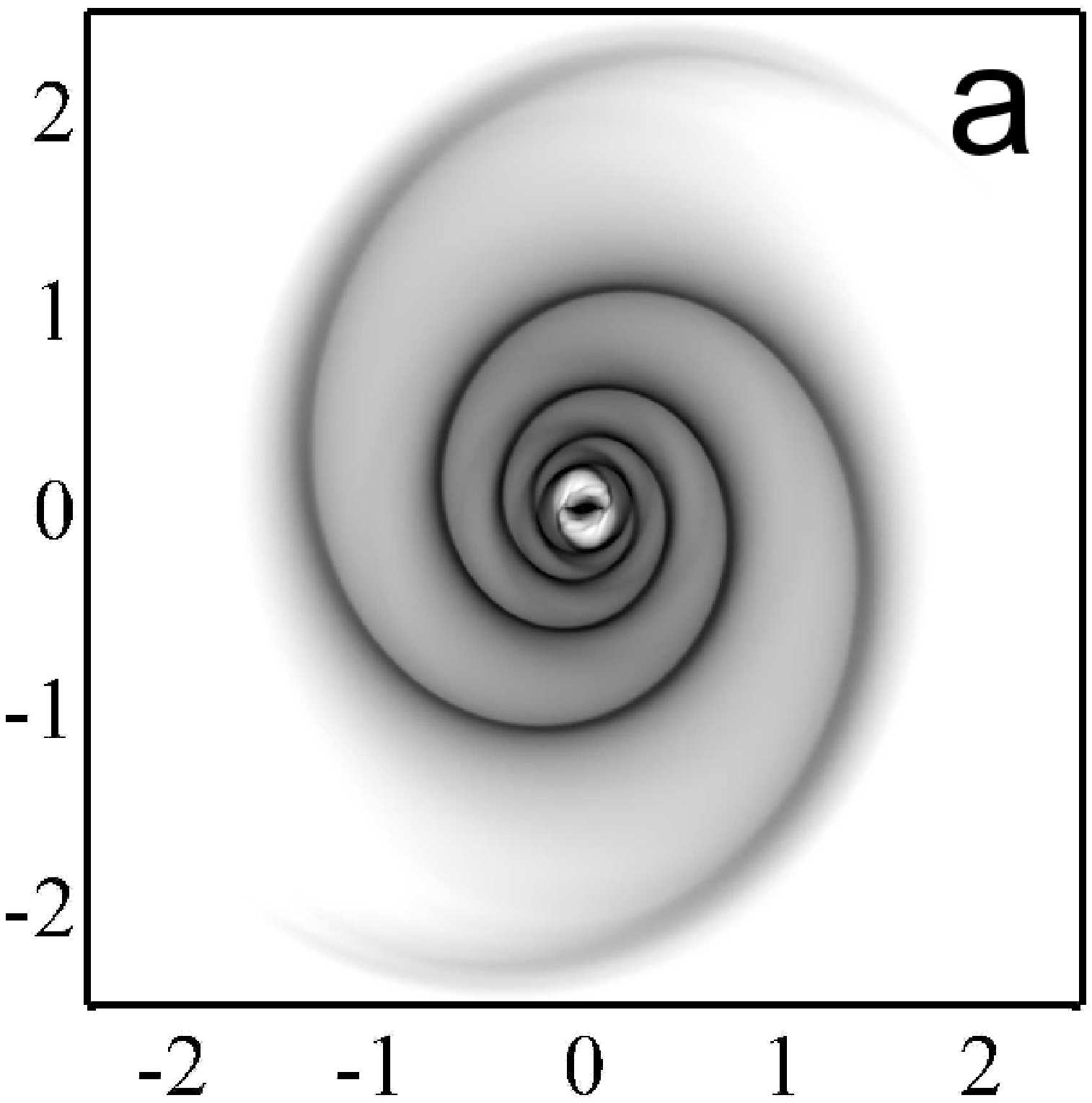}
\includegraphics[width=0.45\hsize]{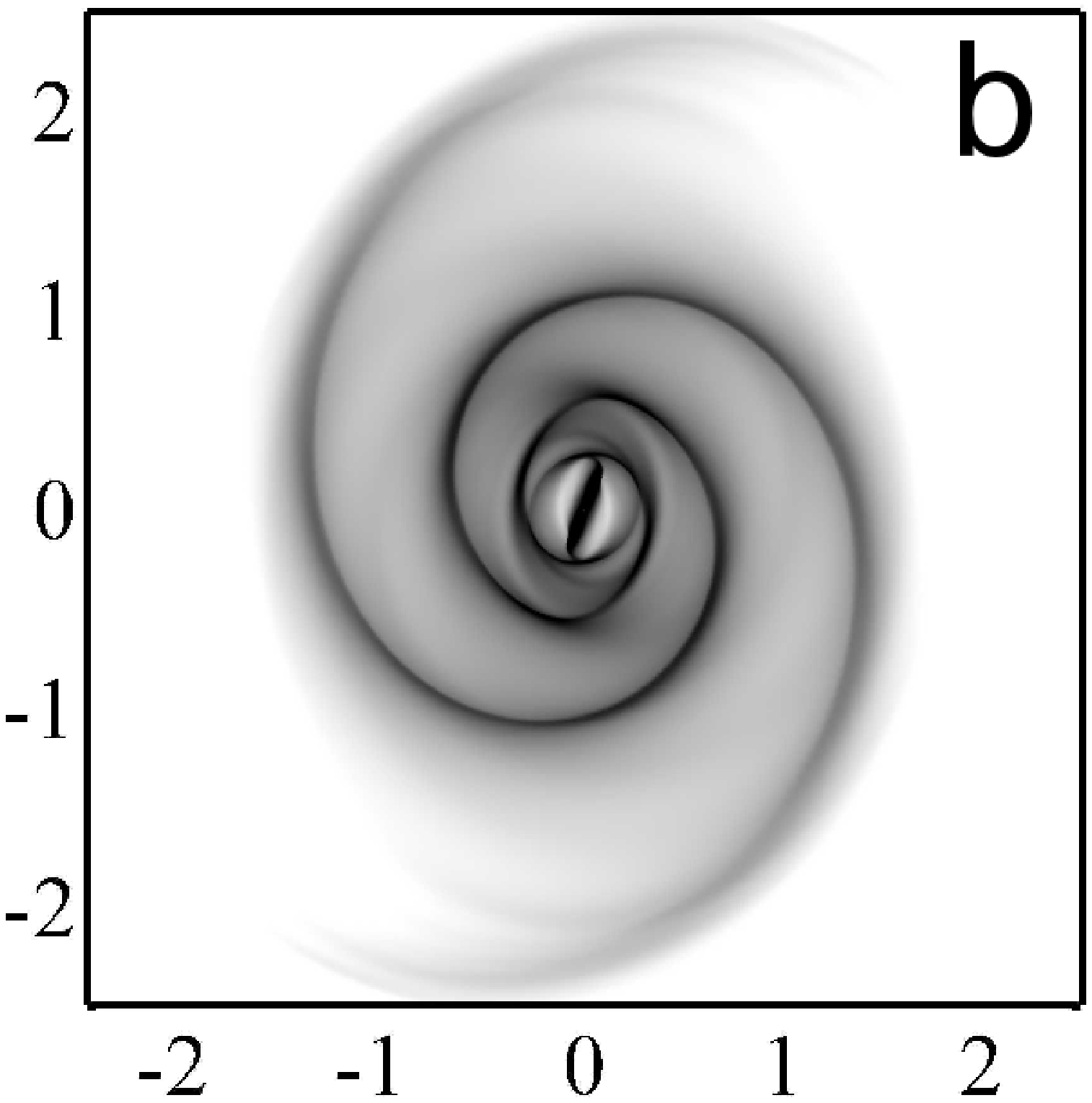}
\includegraphics[width=0.45\hsize]{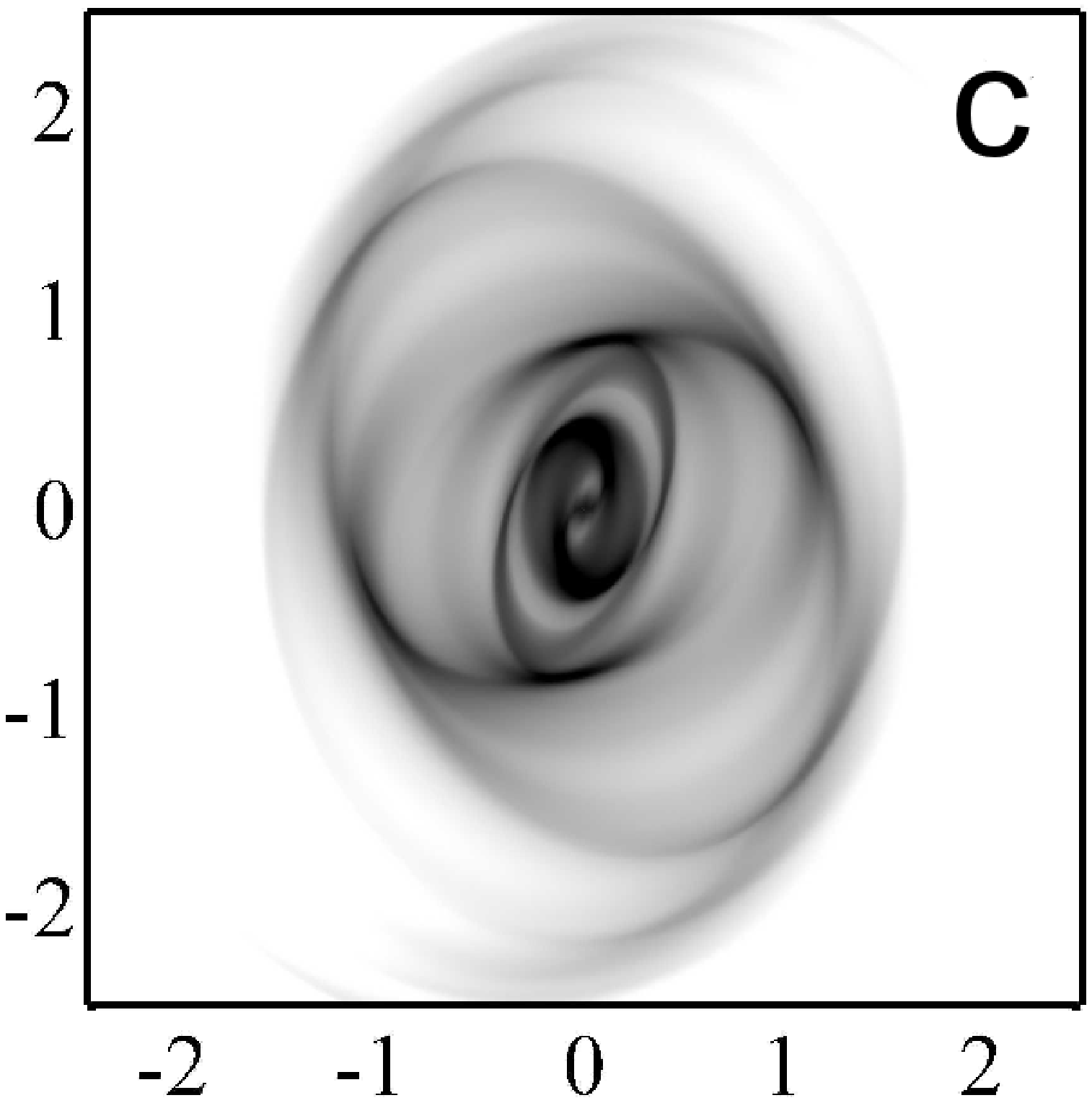}
\captionstyle{normal}
\caption {Spiral patterns with various geometries in the
central region at the same time for ${\cal M}_0=10$ and $\varepsilon=0.1$ ,
in the models with (a) $d=0.02$, (b) $0.1$, and (c) $0.3$.
 }\label{fig-10-spiral-Theta}
\end{figure}
\newpage
%
\begin{figure}
\setcaptionmargin{5mm}
\onelinecaptionsfalse
\begin{tabular}{cc}
\includegraphics[width=0.45\hsize]{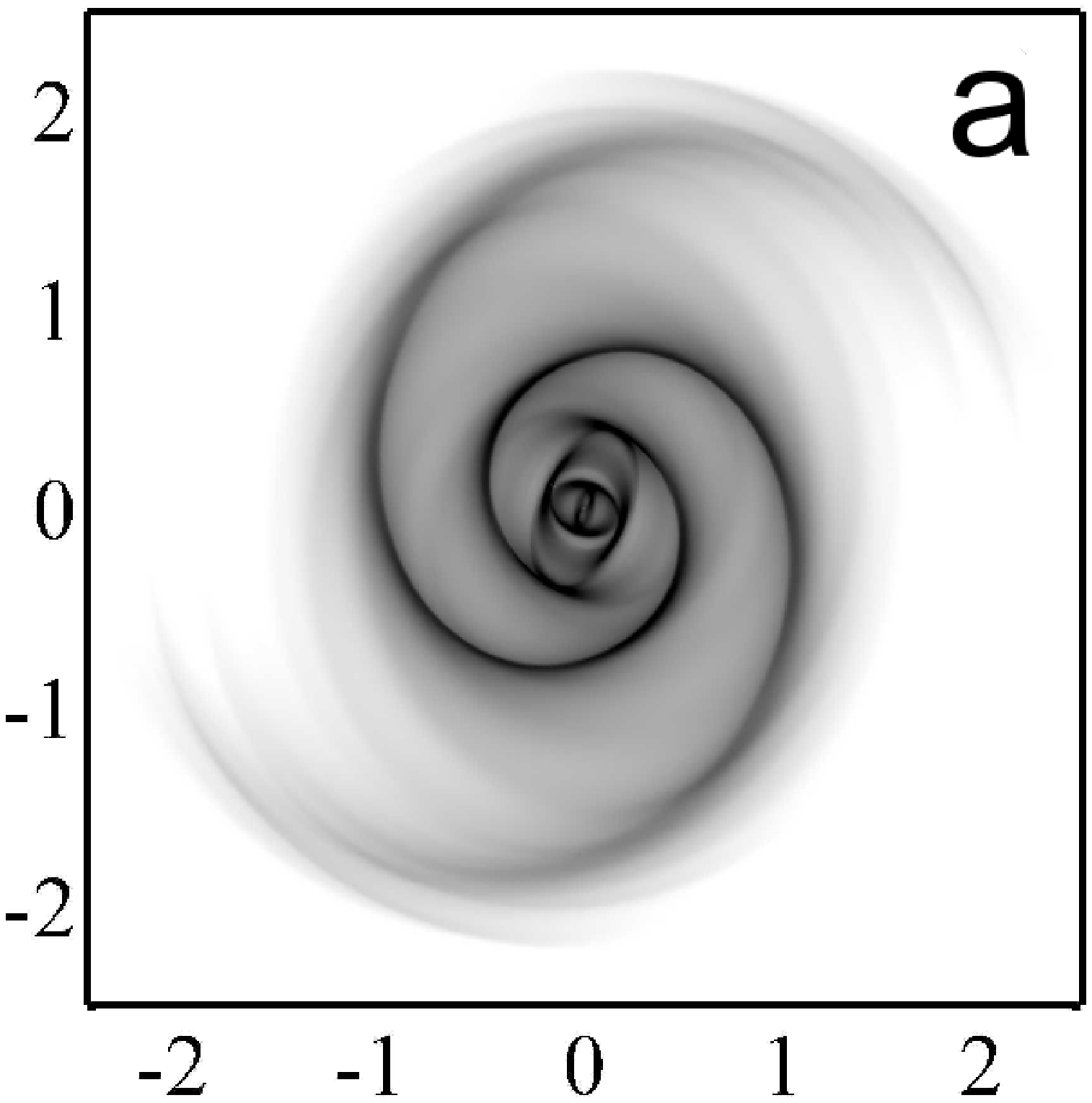} &
\includegraphics[width=0.45\hsize]{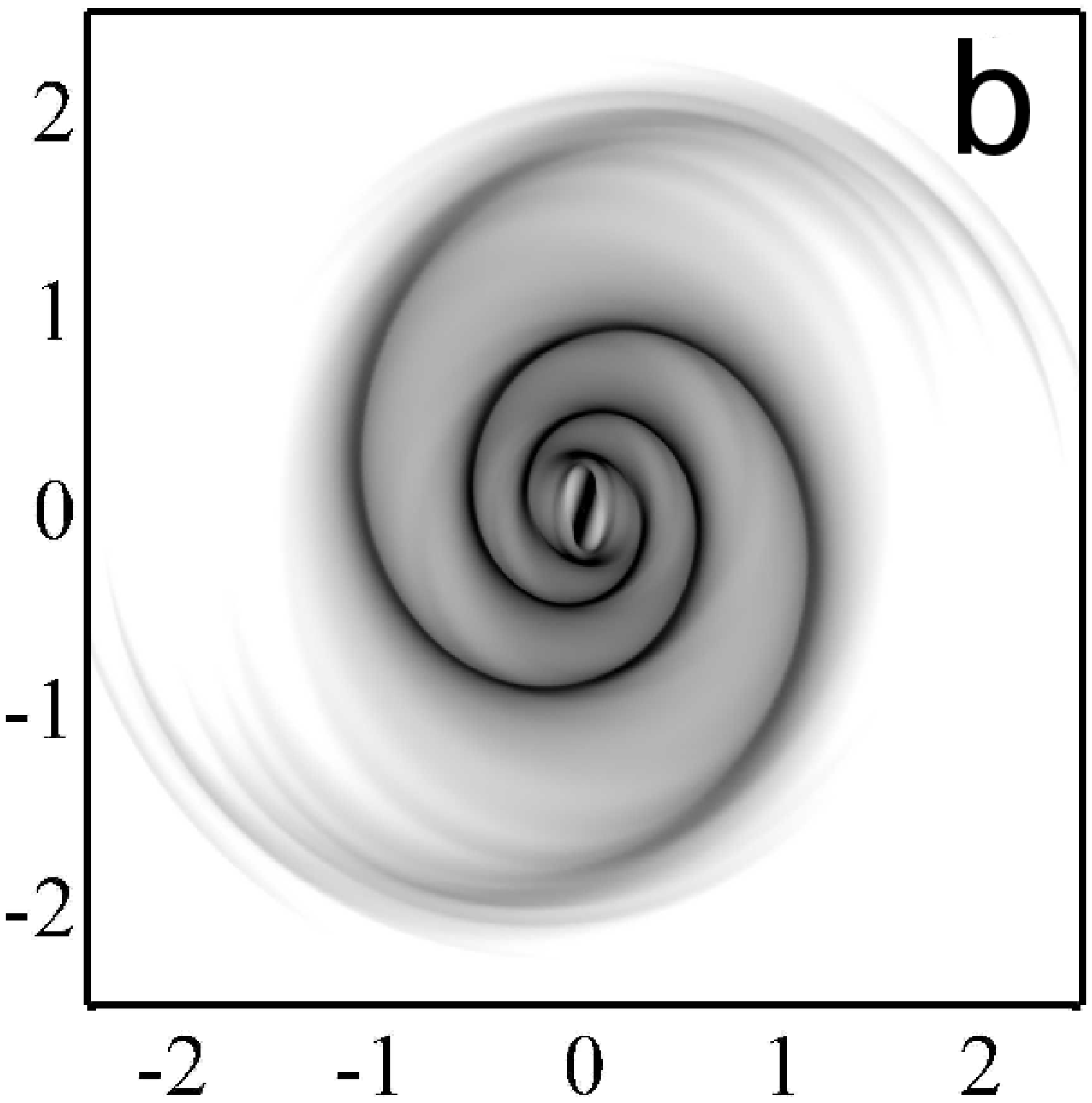} \\
\includegraphics[width=0.45\hsize]{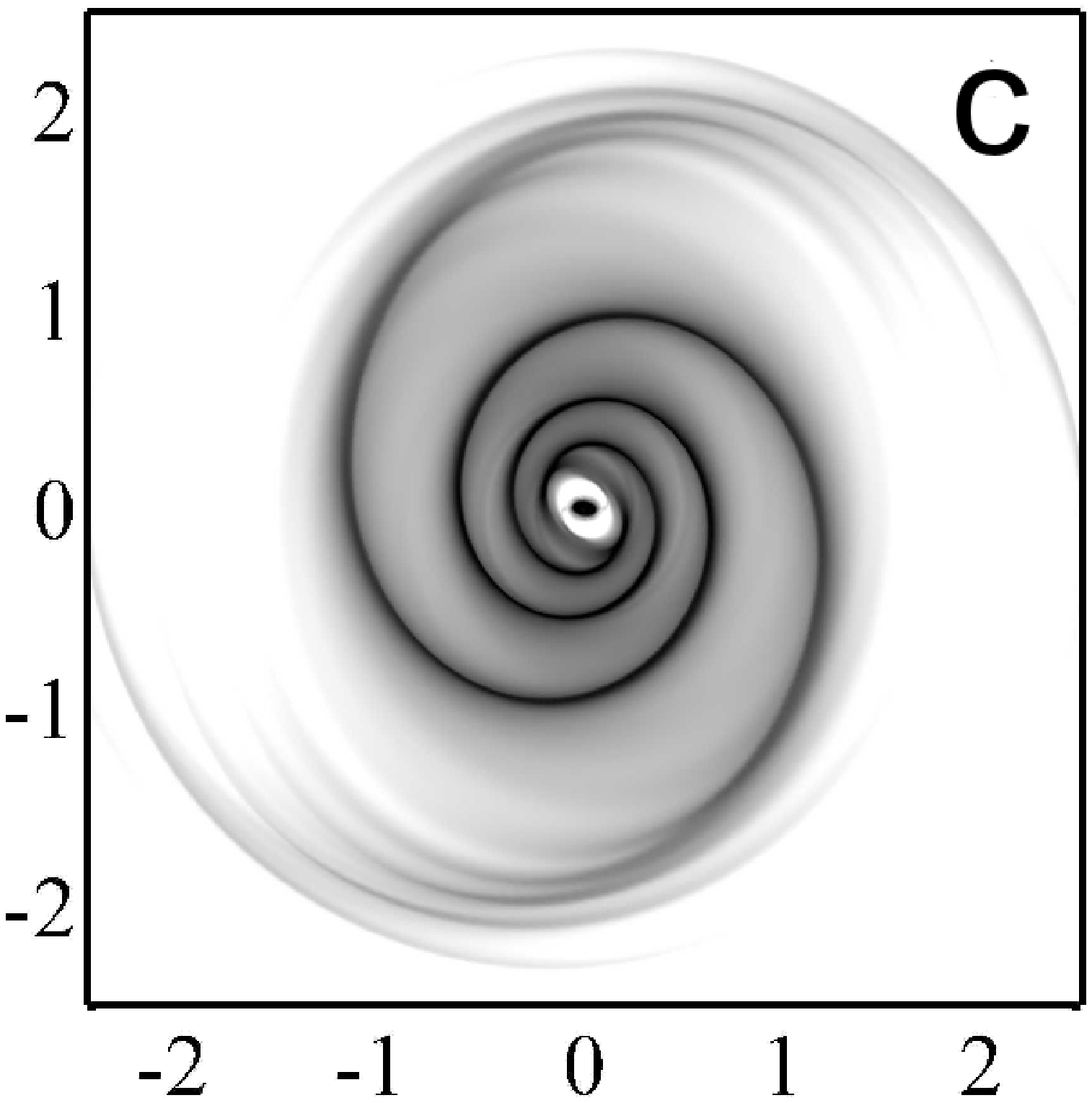} &
\includegraphics[width=0.45\hsize]{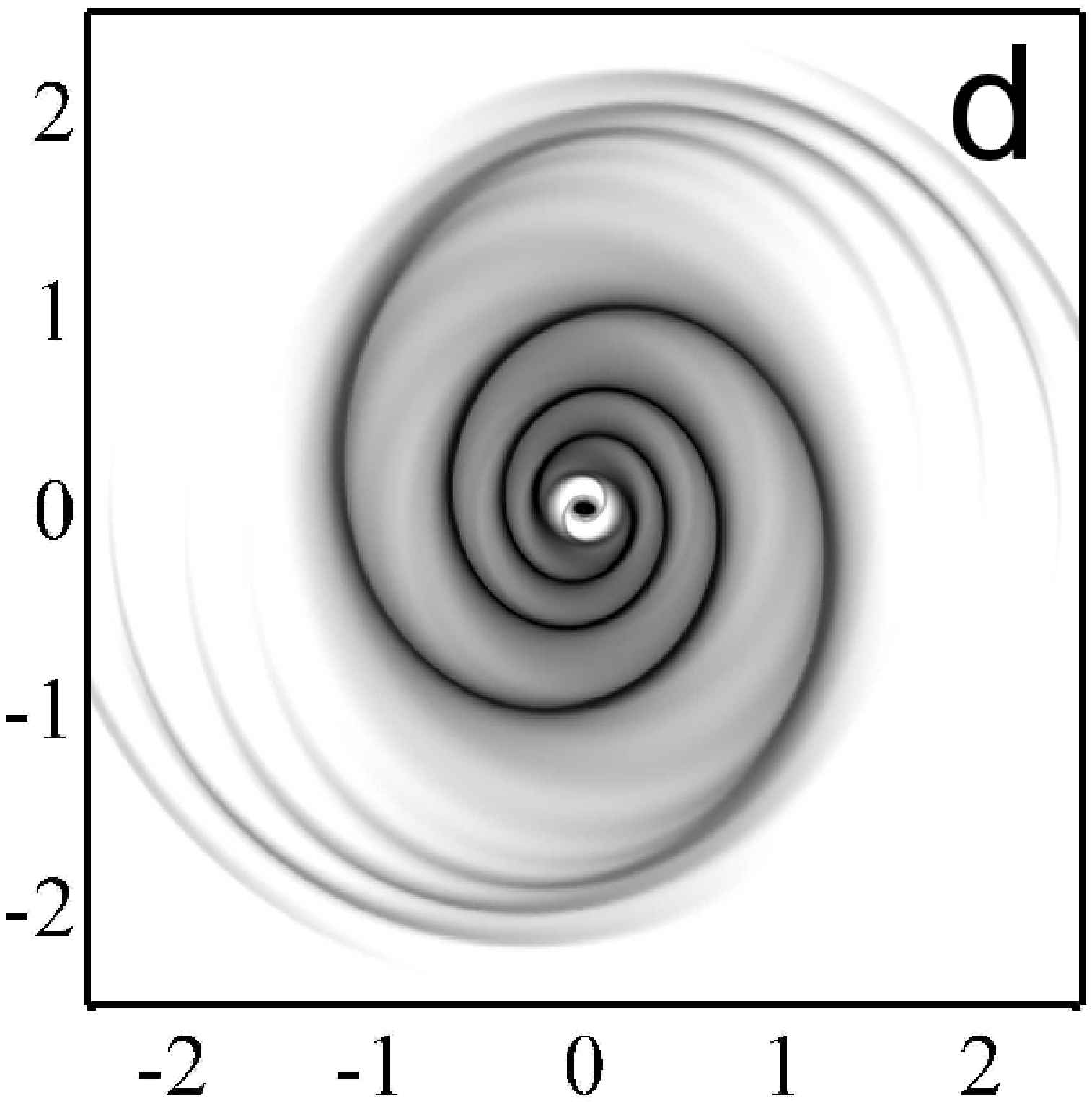}
\end{tabular}
\captionstyle{normal}
\caption {Models with Mach numbers ${\cal M}_0=\text{ (a) }10, \text{ (b) }15, \text{ (c) } 20, \text{  and}\text{ (d) }30$ with $\varepsilon=0.1$ and $d=0.1$, all at the same time. }\label{fig-10-spiral-Mah}
\end{figure}
\newpage
%
\begin{figure}
\setcaptionmargin{5mm}
\onelinecaptionsfalse
\includegraphics[width=0.7\hsize]{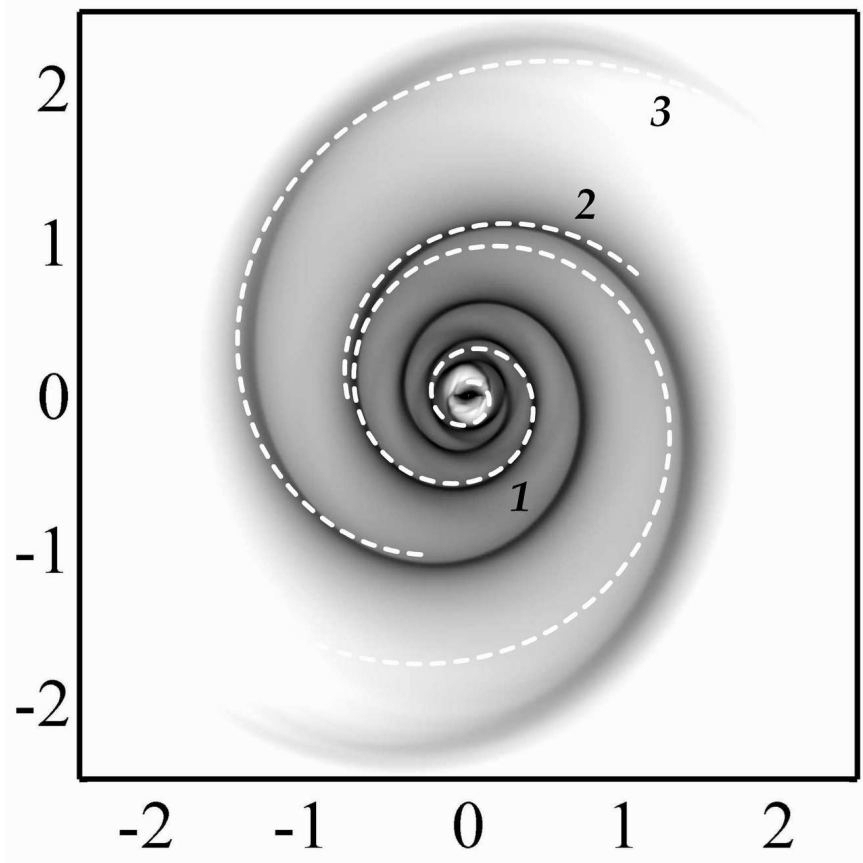}
\captionstyle{normal}
\caption {Geometry of spirals in the model with ${\cal M}_0=10$, $\varepsilon_h=0.1$, and $d=0.02$. Superimposed on the distribution of the
logarithm of the surface density are curves showing sections of logarithmic spirals with various pitch angles: \textit{1} $i=10^\circ$, \textit{2} $i=12^\circ$, \textit{3} $i=14^\circ$. }\label{fig-angle-i}
\end{figure}
\newpage
%
\begin{figure}
\setcaptionmargin{5mm}
\onelinecaptionsfalse
   \includegraphics[width=0.7\hsize]{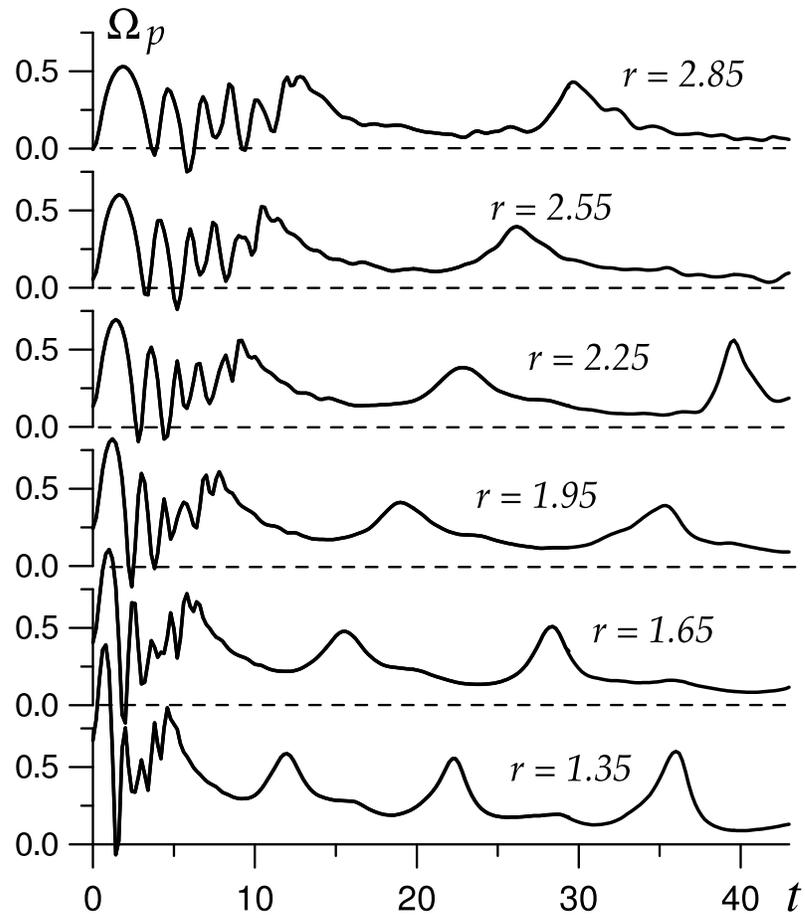}
\captionstyle{normal}
\caption {Dependences $\Omega_p(t)$ for the two-armed spiral for disk rings situated at different distances from the center~$r$.}\label{fig-10-spiral-rot}
\end{figure}
\newpage
%
\begin{figure}
\setcaptionmargin{5mm}
\onelinecaptionsfalse
 \includegraphics[width=0.5\hsize]{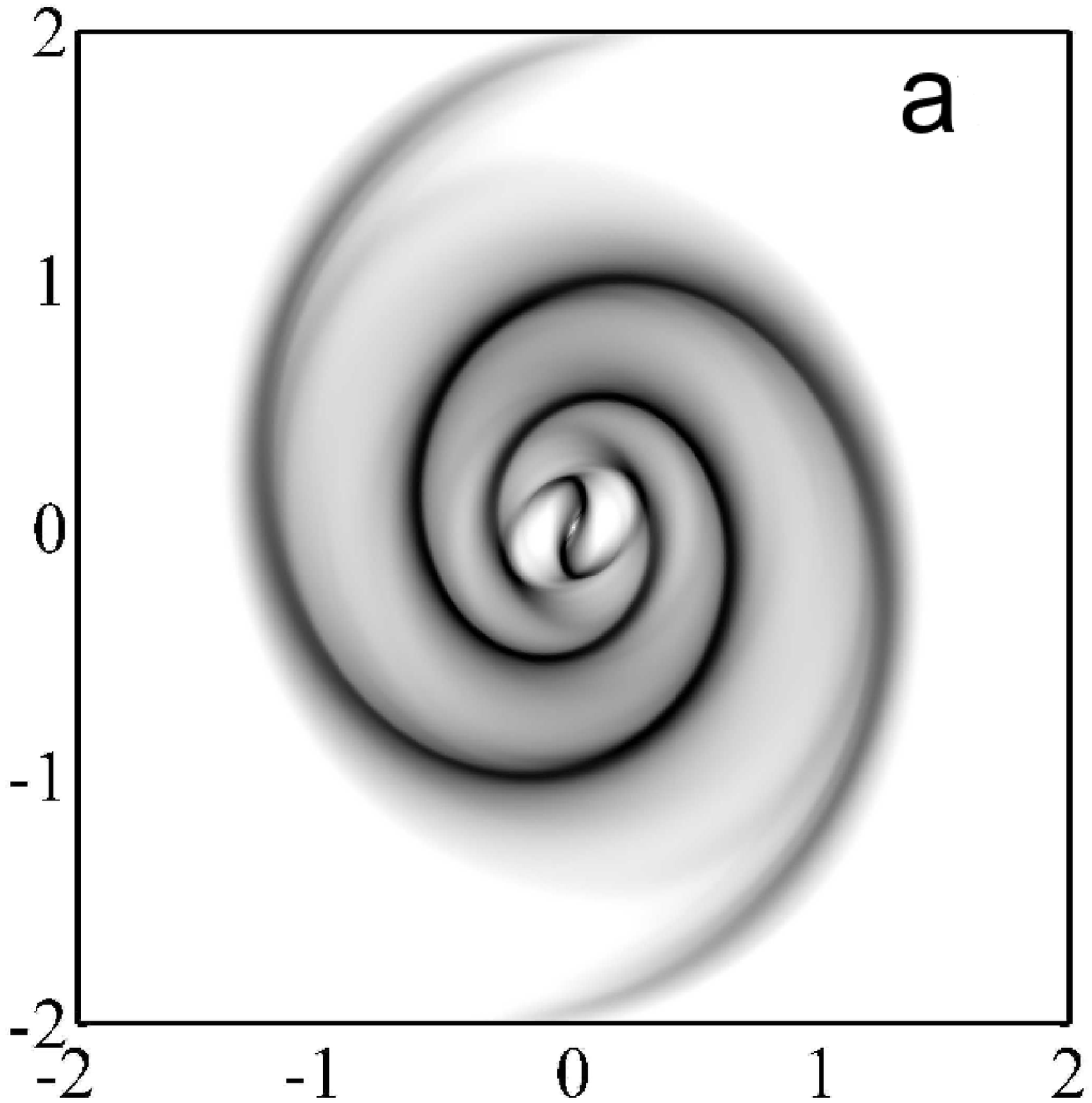}
 \includegraphics[width=0.55\hsize]{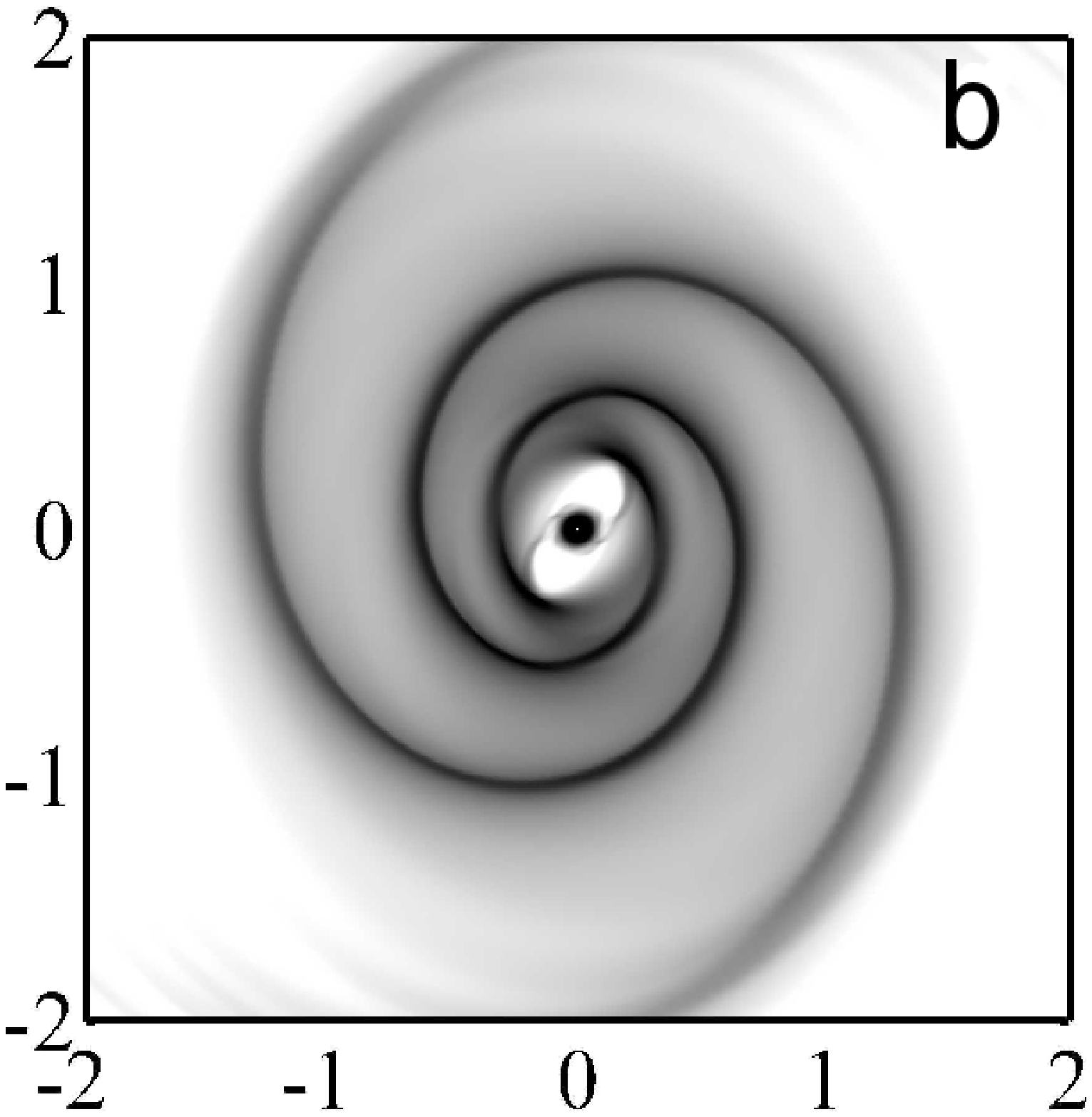}
 \includegraphics[width=0.6\hsize]{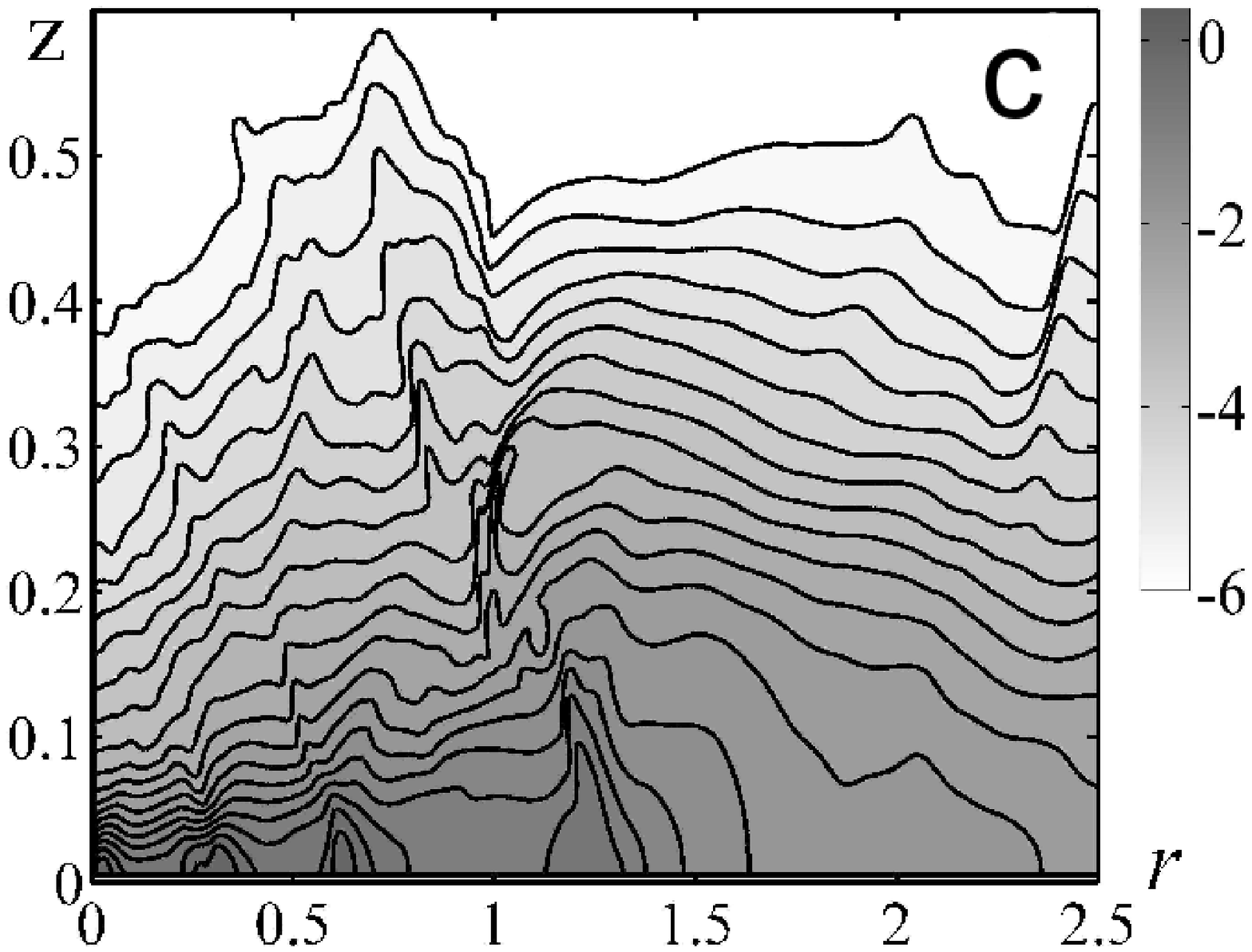}
\captionstyle{normal}
\caption {Distributions of the surface density in the (a) 3D and (b) 2D models in the TVD computations for $\varepsilon_h=0.1$, ${\cal M}_0=15$, and $d=0.125$, and (c) the vertical structure of the volume density in the $(x,0,z)$ plane for the azimuth angle $\varphi=0$. }\label{fig-10-spiral-3D}
\end{figure}
\newpage
%
\begin{figure}
\setcaptionmargin{5mm}
\onelinecaptionsfalse
            \includegraphics[width=0.45\hsize]{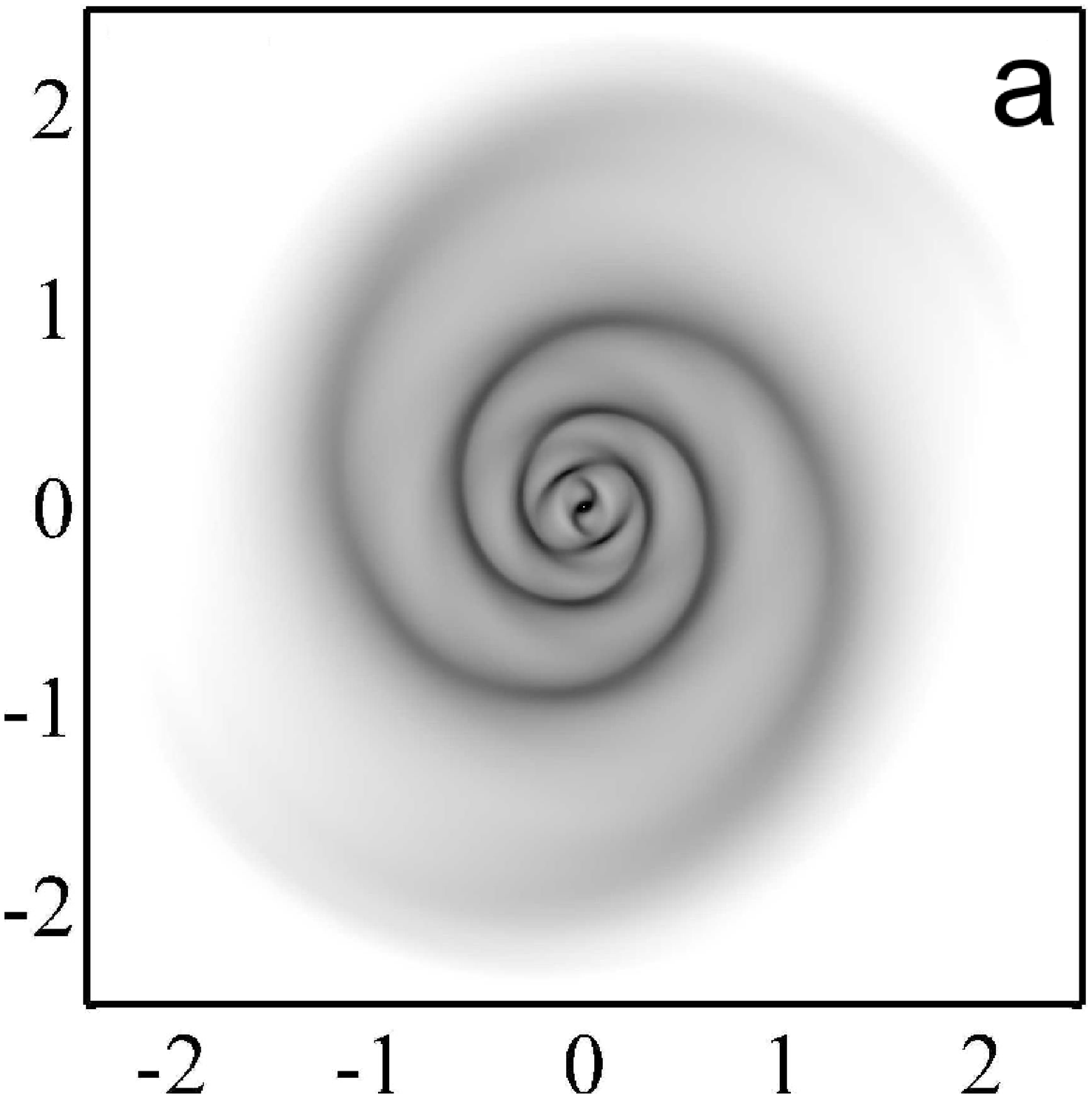}
            \includegraphics[width=0.45\hsize]{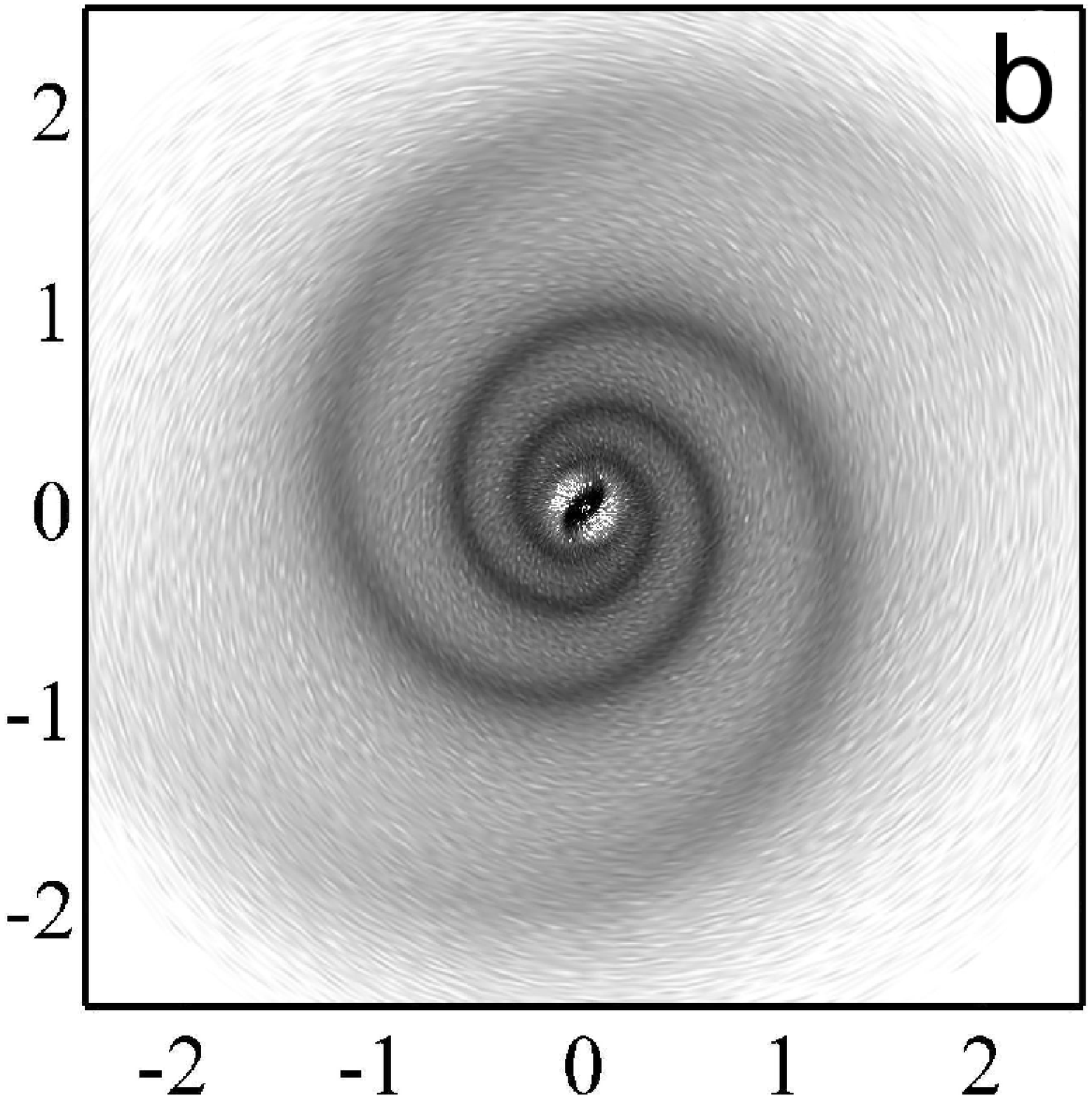}
\captionstyle{normal}
\caption {Example of a comparison of the spiral structure of the (a) 2D TVD model and (b) the 3D SPH computation. }\label{fig-10-spiral-3D-SPH}
\end{figure}
\newpage

%
\begin{figure}
\setcaptionmargin{5mm}
\onelinecaptionsfalse
             \includegraphics[width=0.7\hsize]{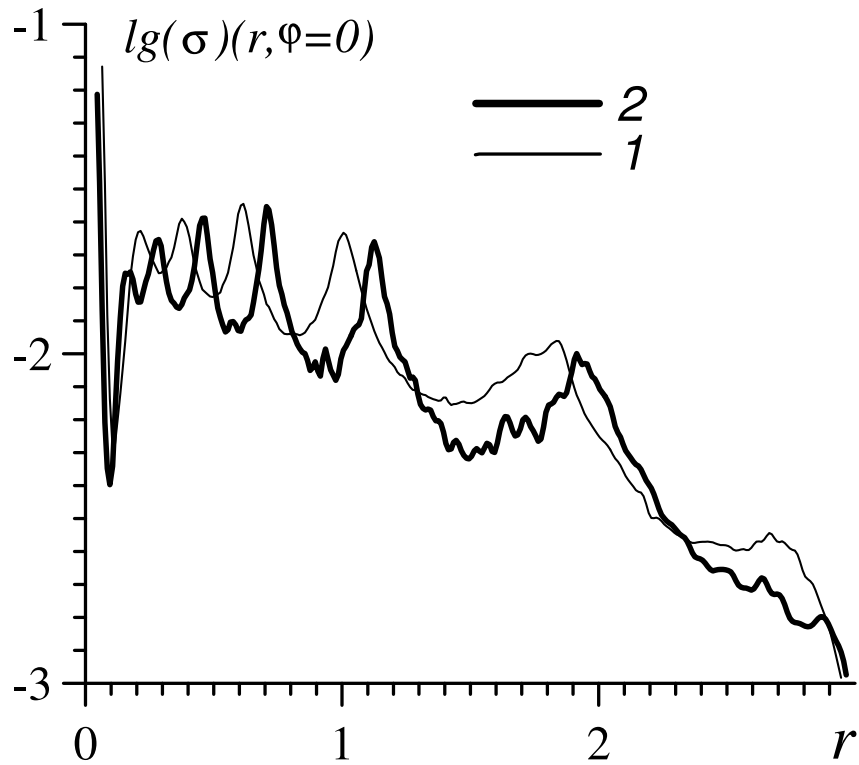}
\captionstyle{normal}
\caption {Radial profiles of the surface density along the ray $\varphi=0$ in the SPH disk model (\textit{1}) without and (\textit{2}) with self-gravity. }\label{fig-10-spiral-self-gr}
\end{figure}

\pagebreak
\pagebreak

\end{document}